\documentclass[modern]{aastex61}
\usepackage{float, bm, graphicx, amsmath, morefloats}
\usepackage{CJK,CJKspace,CJKpunct}
\usepackage{color}

\newcommand{\code}[1]{\texttt{\detokenize{#1}}}
\newcommand{\foreign}[1]{\textsl{#1}}
\newcommand{\etal}{\foreign{et al.}}

\newcommand{\jy}{\mbox{$\rm Jy$}}
\newcommand{\mjy}{\mbox{$\rm mJy$}}
\newcommand{\ujy}{\mbox{$\rm \mu Jy$}}

\newcommand{\days}{\mbox{$\rm days$}}

\newcommand{\pyr}{\mbox{$\rm yr^{-1}$}}
\newcommand{\second}{\mbox{$\rm s$}}
\newcommand{\millisecond}{\mbox{$\rm ms$}}
\newcommand{\psec}{\mbox{$\rm s^{-1}$}}

\newcommand{\erg}{\mbox{$\rm erg$}}
\newcommand{\kev}{\mbox{$\rm keV$}}

\newcommand{\mpc}{\mbox{$\rm Mpc$}}
\newcommand{\kpc}{\mbox{$\rm kpc$}}
\newcommand{\pc}{\mbox{$\rm pc$}}
\newcommand{\cm}{\mbox{$\rm cm$}}
\newcommand{\km}{\mbox{$\rm km$}}
\newcommand{\au}{\mbox{$\rm AU$}}

\newcommand{\ghz}{\mbox{$\rm GHz$}}
\newcommand{\mhz}{\mbox{$\rm MHz$}}

\newcommand{\phz}{\mbox{$\rm Hz^{-1}$}}

\newcommand{\swift}{{\em Swift}}
\newcommand{\nustar}{{\em NuSTAR}}

\newcommand{\pcm}{\mbox{$\rm cm^{-1}$}}
\newcommand{\pcmsq}{\mbox{$\rm cm^{-2}$}}
\newcommand{\cmcub}{\mbox{$\rm cm^{3}$}}
\newcommand{\pcmcub}{\mbox{$\rm cm^{-3}$}}
\newcommand{\pct}{\mbox{$\rm ct^{-1}$}}
\newcommand{\gauss}{\mbox{$\rm G$}}

\newcommand{\gram}{\mbox{$\rm g$}}
\newcommand{\kelvin}{\mbox{$\rm K$}}

\newcommand{\mas}{\mbox{$\rm \mu as$}}

\newcommand{\lum}{\mbox{$\rm erg\,s^{-1}$}}

\newcommand{\radius}{\mbox{$\rm 7 \times 10^{15}\,\cm$}}
\newcommand{\bfield}{\mbox{$\rm 6\,\gauss$}}
\newcommand{\energy}{\mbox{$\rm 4 \times 10^{48}\,\erg$}}

\newcommand{\velocity}{\mbox{$\rm 0.13c$}}
\newcommand{\density}{\mbox{$\rm 3 \times 10^{5}\,\pcmcub$}}
\newcommand{\nupeak}{\mbox{$\rm 100\,GHz$}}
\newcommand{\fpeak}{\mbox{$\rm 94\,mJy$}}
\newcommand{\nucool}{\mbox{$\rm 2\,GHz$}}

\begin{document}

\title{AT2018cow: a luminous millimeter transient}

\author{Anna Y.\ Q.\ ~Ho}
\affiliation{Cahill Center for Astrophysics, 
California Institute of Technology, MC 249-17, 
1200 E California Boulevard, Pasadena, CA, 91125, USA}

\author{E.\ Sterl ~Phinney}
\affiliation{Theoretical Astrophysics, MC 350-17, California Institute of Technology,
Pasadena, CA 91125, USA}

\author{Vikram~Ravi}
\affiliation{Cahill Center for Astrophysics, 
California Institute of Technology, MC 249-17, 
1200 E California Boulevard, Pasadena, CA, 91125, USA}
\affiliation{Harvard-Smithsonian Center for Astrophysics,
60 Garden Street, Cambridge, MA 02138, USA}

\author{S.\ R.\ ~Kulkarni}
\affiliation{Cahill Center for Astrophysics, 
California Institute of Technology, MC 249-17, 
1200 E California Boulevard, Pasadena, CA, 91125, USA}

\author{Glen Petitpas}
\affiliation{Harvard-Smithsonian Center for Astrophysics,
60 Garden Street, Cambridge, MA 02138, USA}

\author{Bjorn Emonts}
\affiliation{National Radio Astronomy Observatory, 520 Edgemont Road, Charlottesville, VA 22903, USA}

\author{V.\ Bhalerao}
\affiliation{Department of Physics, Indian Institute of Technology Bombay, Mumbai 400076, India}

\author{Ray Blundell}
\affiliation{Harvard-Smithsonian Center for Astrophysics,
60 Garden Street, Cambridge, MA 02138, USA}

\author{S.\ Bradley ~Cenko}
\affiliation{Astrophysics Science Division, NASA Goddard Space Flight Center, Mail Code 661, Greenbelt, MD 20771, USA}
\affiliation{Joint Space-Science Institute, University of Maryland, College Park, MD 20742, USA}

\author{Dougal Dobie}
\affiliation{Sydney Institute for Astronomy, School of Physics, University of Sydney, Sydney, New South Wales 2006, Australia}
\affiliation{ATNF, CSIRO Astronomy and Space Science, PO Box 76, Epping, New South Wales 1710, Australia}

\author{Ryan Howie}
\affiliation{Harvard-Smithsonian Center for Astrophysics,
60 Garden Street, Cambridge, MA 02138, USA}

\author{Nikita Kamraj}
\affiliation{Cahill Center for Astrophysics, 
California Institute of Technology, MC 249-17, 
1200 E California Boulevard, Pasadena, CA, 91125, USA}

\author{Mansi M.\ ~Kasliwal}
\affiliation{Cahill Center for Astrophysics, 
California Institute of Technology, MC 249-17, 
1200 E California Boulevard, Pasadena, CA, 91125, USA}

\author{Tara Murphy}
\affiliation{Sydney Institute for Astronomy, School of Physics, University of Sydney, Sydney, New South Wales 2006, Australia}

\author{Daniel A.\ ~Perley}
\affiliation{Astrophysics Research Institute, Liverpool John Moores University, IC2, Liverpool Science Park, 146 Brownlow Hill, Liverpool L3 5RF, UK}

\author{T. K. Sridharan}
\affiliation{Harvard-Smithsonian Center for Astrophysics,
60 Garden Street, Cambridge, MA 02138, USA}

\author{Ilsang Yoon}
\affiliation{National Radio Astronomy Observatory, 520 Edgemont Road, Charlottesville, VA 22903, USA}

\begin{abstract}

We present detailed submillimeter- through centimeter-wave observations of the extraordinary extragalactic transient AT2018cow.
The apparent characteristics --- the high radio luminosity,
the rise and long-lived emission plateau
at millimeter bands, and the sub-relativistic velocity --- have no precedent.
A basic interpretation of the data suggests $E_k \gtrsim \energy$
coupled to a fast but sub-relativistic ($v \approx \velocity$) shock in a dense ($n_e \approx \density$) medium.  We find that the X-ray emission is not naturally explained by an extension of the radio--submm synchrotron spectrum, nor by inverse Compton scattering of the dominant blackbody UVOIR photons by energetic electrons within the forward shock.
By $\Delta t \approx20\,\days$,
the X-ray emission shows spectral softening
and erratic inter-day variability.
Taken together, we are led to invoke
an additional source of X-ray emission:
the central engine of the event.
Regardless of the nature of this central engine, this source heralds a new class of energetic transients shocking a dense medium, which at early times are most readily observed at millimeter wavelengths.

\end{abstract}

\section{Introduction}
\label{sec:introduction}

\subsection{The transient millimeter sky}

Although the sky is regularly monitored
across many bands of the electromagnetic spectrum
(as well as in gravitational waves and energetic particles)
the dynamic sky at millimeter to sub-millimeter wavelengths
(0.1--10\,mm) remains poorly explored.
There has only been one blind transient survey
specific to the millimeter band\footnote{The authors searched for transient sources at 90\,\ghz\ and 150\,\ghz. They found a single candidate event, which intriguingly showed linear polarization.} \citep{Whitehorn2016};
millimeter facilities are usually only triggered after an initial discovery at another wavelength.
Even when targeting known transients, the success rate for detection is low, and to date
only a few extragalactic transients\footnote{Here we use ``transient'' as distinct from ``variable'': millimeter observations are used to study variability in protostars (e.g. \citealt{Herczeg2017}) and more commonly for active galactic nuclei (e.g. \citealt{Dent1983})} have well-sampled, multifrequency light curves.
This sample includes supernovae (SNe; \citealt{Weiler2007,Horesh2013}), tidal-disruption events
(TDEs; \citealt{Zauderer2011,Yuan2016}) and
gamma-ray bursts (GRBs; \citealt{deUgarte2012,Laskar2013,Perley2014,Urata2014}).
There have also been millimeter detections of galactic transient sources, primarily stellar flares (e.g. \citealt{Bower2003}, \citealt{Fender2015}). 

The paucity of millimeter transient studies
can be attributed in part to
costly receiver and electronics systems and
the need for excellent weather conditions,
but it also reflects challenges intrinsic to millimeter-wave transients themselves:
most known classes are either too dim (SNe, most TDEs) to detect unless they are very nearby, or too short-lived (GRBs) to detect without very rapid reaction times ($<$1 day, and even in these circumstances the emission may only be apparent from low-density environments; \citealt{Laskar2013}).

An evolving technical landscape,
together with rapid follow-up enabled by high-cadence optical surveys,
present new opportunities for millimeter transient astronomy.
Lower-noise receivers and ultra-wide bandwidth capability
have greatly increased the sensitivity of sub-mm facilities (e.g.\ the  Submillimeter Array or SMA; \citealt{Ho2004}), and the Atacama Large Millimeter Array (ALMA), a flagship facility, recently began operations.
Optical surveys are discovering new and unexpected classes of transient events whose millimeter properties are unknown --- and possibly different from previously-known types --- motivating renewed follow-up efforts.

\subsection{AT2018cow}

AT2018cow was discovered on 2018 June 16 UT as an optical transient \citep{Smartt2018,Prentice2018}
by the Asteroid Terrestrial-impact Last Alert System (ATLAS; \citealt{Tonry2018}).
It attracted immediate attention because of its fast rise time ($t_\mathrm{peak} \lesssim 3\,$ days),
which was established by earlier non-detections  \citep{Fremling2018,Prentice2018},
together with its high optical luminosity ($M_\mathrm{peak} \sim -20$) and its close proximity ($d=60\,$Mpc).

UVOIR observations \citep{Prentice2018,Perley2018}
revealed unprecedented photometric and spectroscopic properties.
Long-lived luminous X-ray emission was detected
with \swift/XRT \citep{RiveraSandoval2018a},
INTEGRAL \citep{Ferrigno2018,Savchenko2018}
and NuSTAR \citep{Margutti2018a,Grefenstette2018}.
Early radio and sub-millimeter detections were reported by NOEMA \citep{deUgarte2018},
JCMT \citep{Smith2018}, AMI \citep{Bright2018}, and by us using the ATCA \citep{Dobie2018a,Dobie2018b}.
The source does not appear to be a GRB,
as no prompt high-energy emission was detected in searches of \swift/BAT \citep{Lien2018}, Fermi/GBM \citep{DalCanton2018}, Fermi/LAT \citep{Kocevski2018},
and AstroSat CZTI \citep{Sharma2018}.

\citet{Perley2018} suggested that AT2018cow is a new member of the class of rapidly rising ($t_{\rm rise} \lesssim 5$d) and luminous ($M_{\rm peak} < -18$) blue transients,
which have typically been found
in archival searches of optical surveys \citep{Drout2014,Pursiainen2018,Rest2018}.
The leading hypothesis for this class was
circumstellar interaction of a supernova \citep{Ofek2010},
but this was difficult to test because most of the events were located at cosmological distances, and not discovered in real-time.
AT2018cow presented the first opportunity to study a member of this class up close and in real time,
but its origin remains mysterious
despite the intense ensuing observational campaign.
Possibilities include failed supernovae and tidal disruption events, but although AT2018cow shares properties with both of these classes, it is clearly not a typical member of either \citep{Prentice2018,Perley2018,Kuin2018}.

Given the unusual nature of the source, we were motivated to undertake high-frequency observations.
We began monitoring AT2018cow with the SMA 
at 230\,\ghz\ and 340\,\ghz\ and carried out 
supporting observations
with the ATCA
from 5\,\ghz\ to 34\,\ghz.
To our surprise the source was very bright and still rising at sub-millimeter wavelengths (and optically thick in the centimeter band) days after the discovery.
Our extensive SMA observations represent the first millimeter observation of a transient in its rise phase.

This finding led us to seek Director's discretionary (DD) time with ALMA at even higher frequencies, which enabled us to resolve the peak of the SED.
A technical highlight of the ALMA observations
was the detection of the source at nearly a terahertz frequency (Band 9).
We present the sub-millimeter, radio, and X-ray observations in Section \ref{sec:obs},
and our modeling of the radio-emitting ejecta in Section \ref{sec:modeling}.
In Section \ref{sec:implications} we put our velocity and energy measurements in the context of other transients (\ref{sec:velocity-energy}),
attribute the high sub-mm luminosity of AT2018cow to the large density of the surrounding medium (\ref{sec:luminosity}), 
and discuss some problems with the synchrotron model parameters  (\ref{sec:breakfreq}).
In Section \ref{sec:xray} we attribute the late-time X-ray emission to a powerful central engine.
We look ahead to the future in Section \ref{sec:outlook}.

\newpage

\section{Observations}
\label{sec:obs}

All observations are measured $\Delta t$ (observer-frame) from the zero-point MJD 58285 (following \citealt{Perley2018}),
which lies between the date of discovery (MJD 58285.441) and the last non-detection (58284.13; \citealt{Prentice2018}).
At $\Delta t=14\,\days$ we find excellent agreement between
the SMA and the ALMA data, showing that the flux scales are consistent.

\subsection{Radio and submillimeter observations}
\label{sec:obsradio}

\subsubsection{The Submillimeter Array (SMA)}
\label{sec:sma}

AT2018cow was regularly observed with the SMA under its Director Discretionary Time/Target of opportunity program.
Observations took place over the period of UT 2018 Jun 21--UT 2018 August 3
($\Delta t\approx5$--$49$\,\days)
in the Compact configuration,
with an additional epoch on UT 2018 August 31
($\Delta t\approx76$\,\days).
All observations contained 6 to 8 antennas and cover a
range of baseline lengths from 16.4\,m to 77\,m.
A majority of these observations were short and were
repeated almost nightly by sharing tuning and calibration data with
other science tracks. The SMA has two receiver sets each with 8\,GHz of
bandwidth in each of two sidebands (32\,GHz total) covering a range of
frequencies from 188--416\,GHz. Each receiver can be tuned independently
to provide dual-band observations.  Additionally the upper and lower
sidebands are separated (center to center) by 16\,GHz allowing up to four
simultaneous frequency measurements. During some observations, the receivers
were tuned to the same local oscillator frequency allowing
the lower and upper sidebands to be averaged together, improving the signal-to-noise ratio. For all
observations, the quasars 1635+381 and 3C\,345 were used as primary phase and
amplitude gain calibrators, respectively, with absolute flux calibration performed by
nightly comparison to Titan, Neptune, or (maser-free) continuum
observations of the emission-line star MWC349a. The quasar 3C\,279 and/or the blazar 3C\,454.3 was used for bandpass
calibration. Data were calibrated in IDL using the MIR package. Additional
analysis and imaging were performed using the MIRIAD package. Given
that the target was a point source, fluxes were derived directly from the
calibrated visibilities, but the results agree well with flux
estimates derived from the CLEANed images when the data quality and
uv-coverage was adequate.

\subsubsection{The Australia Telescope Compact Array (ATCA)}
\label{sec:atca}

We obtained six epochs of centimeter-wavelength observations with the Australia Telescope Compact Array (ATCA; \citealt{Frater92}).
During the first three epochs, the six 22-m dishes were arranged in an east-west 1.5A configuration, with baselines ranging from 153\,m to 4469\,m. During the latter three epochs, five of the six dishes were moved to a compact H75\footnote{https://www.narrabri.atnf.csiro.au/operations/array\_configurations/configurations.html} configuration, occupying a cardinally oriented `T' with baselines ranging from 31\,m to 89\,m. Full-Stokes data were recorded with the Compact Array Broadband Backend \citep{Wilson2011} in a standard continuum \textit{CFB\,1M} setup, simultaneously providing two 2.048\,GHz bands each with 2048 channels. Observations were obtained with center frequencies of 5.5\,GHz \& 9\,GHz, 16.7\,GHz \& 21.2\,GHz, and 33\,GHz \& 35\,GHz, with data in the latter two bands typically being averaged to form a band centered at 34\,GHz. The flux density scale was set using observations of the ATCA flux standard PKS\,1934$-$638. For observations below 33\,GHz, PKS\,1934$-$638 was also used to calibrate the complex time-independent bandpasses, and regular observations of the compact quasar PKS\,1607$+$268 were used to calibrate the time-variable complex gains. For the higher-frequency observations, a brighter source (3C\,279) was used for bandpass calibration (except for epochs 1 and 4), and the compact quasar 4C\,10.45 was used for gain calibration. In the H75 configuration, we only report results from observations at 34\,GHz, from baselines not subject to antenna shadowing. For all 34\,GHz observations, data obtained with the sixth antenna located 4500\,m from the center of the array were discarded because of the difficulty of tracking the differential atmospheric phase over the long baselines to this antenna. The weather was good for all observations, with negligible wind and $<500$\,$\mu$m of rms atmospheric path-length variations \citep{Middelberg2006}.

The data were reduced and calibrated using standard techniques implemented in the MIRIAD software \citep{Sault1995}. To search for unresolved emission at the position of AT2018cow, we made multi-frequency synthesis images with uniform weighting. Single rounds of self-calibration over 5--10\,min intervals were found to improve the image quality in all bands. For data at 5.5\,GHz and 9\,GHz, point-source models of all strong unresolved field sources were used for self-calibration. For data at the higher frequencies, self-calibration was performed using a point-source model for AT2018cow itself, as no other sources were detected within the primary beams, and AT2018cow was detected with a sufficient signal-to-noise ratio. We report flux densities derived by fitting point-source models to the final images using the MIRIAD task \textit{imfit}. 

\subsubsection{The Atacama Large Millimeter/submillimeter Array (ALMA)}
\label{sec:alma}

AT2018cow was observed with ALMA as part of DD time during Cycle 5 using Bands 3, 4, 7, 8, and 9. Observations were performed on 30 June 2018 ($\Delta t\approx14\,\days$; Bands 7 and 8), 08 July 2018 ($\Delta t\approx22\,\days$; Bands 3 and 4) and on 10 July 2018 ($\Delta t\approx23\,\days$; Band 9).\footnote{Band 9 observations were also performed on 09 July 2018, but these data were of too poor quality to use as a result of weather conditions.}

The ALMA 12-m antenna array was in its most compact C43-1 configuration, with 46--48 working antennas and baselines ranging from 12--312\,m. The on-source integration time was 6--8\,min for Bands 3--8, and 40\,min for Band 9. 
The Band 3--8 observations used two-sideband (2SB) receivers with 4\,\ghz\ bandwidth each centered on 91.5 and 103.5\,\ghz\ (Band 3), 138 and 150\,GHz (Band 4), 337.5 and 349.5\,GHz (Band 7), 399 and 411\,GHz (Band 8).
The Band 9 observations used double-sideband (DSB) receivers with 8\,\ghz\ bandwidth (2 times larger than that 
for the Band 3--8 observation, by using 90 degree Walsh phase switching) centered on 663 and 679\,\ghz.
All calibration and imaging was done with the Common Astronomical Software Applications (CASA; \citealt{mcmullin2007}). The data in Bands 3--8 were calibrated with the standard ALMA pipeline, using J1540+1447, J1606+1814 or J1619+2247 to calibrate the complex gains, and using J1337$-$1257 (Band 7), J1550+0527 (Band 3/4) or J1517$-$2422 (Band 8) to calibrate the bandpass response and apply an absolute flux scale. Band 9 observations were delivered following manual calibration by the North American ALMA Science Center, using J1540+1447 for gain calibration, and J1517$-$2422 for bandpass- and flux-calibration. We subsequently applied a phase-only self-calibration using the target source (for Bands 3--8), performed a deconvolution, imaged the data, and flux-corrected for the response of the primary beam. AT2018cow is unresolved in our ALMA data, with a synthesized beam that ranges from $3.3\arcsec\,\times\,2.5$\arcsec\ 
($\mathrm{PA}=29^{\circ}$) in Band 3 to $0.50\arcsec\,\times\,0.36$\arcsec\ ($\mathrm{PA}=-46^{\circ}$) in Band 9. The signal-to-noise ratio in the resulting images ranges from $\sim$500 in Bands 3 and 4 to $\sim$80 in Band 9.
Details about the ALMA Band 9 data reduction
can be found in Appendix \ref{sec:appendix-band9}.

\startlongtable 
\begin{deluxetable}{lrrr} 
\tablecaption{Flux-density measurements for AT2018cow.
Time of detection used is mean UT of observation.
SMA measurements have formal uncertainties shown,         which are appropriate for in-band measurements on a given night.         However, for night-to-night comparisons, true errors are dominated by         systematics and are roughly 10\%--15\% unless indicated otherwise.         ALMA measurements have roughly 5\% uncertainties in Bands 3 and 4,         10\% uncertainties in Bands 7 and 8, and a 20\% uncertainty in Band 9.        ATCA measurements have formal errors listed,         but also have systematic uncertainties of roughly 10\%.\label{tab:flux}} 
\tablewidth{0pt} 
\tablehead{ \colhead{$\Delta t$ (d)} & \colhead{Facility} & \colhead{Frequency (GHz)} & \colhead{Flux Density (mJy)} } 
\tabletypesize{\scriptsize} 
\startdata 
5.39 & SMA & 215.5 & $15.14 \pm 0.56$ \\ 
5.39 & SMA & 231.5 & $16.19 \pm 0.65$ \\ 
6.31 & SMA & 215.5 & $31.17 \pm 0.87$ \\ 
6.31 & SMA & 231.5 & $31.36 \pm 0.97$ \\ 
7.37 & SMA & 215.5 & $40.19 \pm 0.56$ \\ 
7.37 & SMA & 231.5 & $41.92 \pm 0.66$ \\ 
7.41 & SMA & 330.8 & $36.39 \pm 2.25$ \\ 
7.41 & SMA & 346.8 & $30.7 \pm 1.99$ \\ 
8.37 & SMA & 215.5 & $41.19 \pm 0.47$ \\ 
8.37 & SMA & 231.5 & $41.44 \pm 0.56$ \\ 
8.38 & SMA & 344.8 & $26.74 \pm 1.42$ \\ 
8.38 & SMA & 360.8 & $22.79 \pm 1.63$ \\ 
9.26 & SMA & 243.3 & $35.21 \pm 0.75$ \\ 
9.26 & SMA & 259.3 & $36.1 \pm 1.0$ \\ 
9.28 & SMA & 341.5 & $22.85 \pm 1.53$ \\ 
9.28 & SMA & 357.5 & $25.84 \pm 2.5$ \\ 
10.26 & SMA & 243.3 & $36.6 \pm 0.81$ \\ 
10.26 & SMA & 259.3 & $31.21 \pm 0.92$ \\ 
10.26 & SMA & 341.5 & $19.49 \pm 1.47$ \\ 
10.26 & SMA & 357.5 & $17.42 \pm 2.8$ \\ 
11.26 & SMA & 243.3 & $22.14 \pm 1.05$ \\ 
11.26 & SMA & 259.3 & $20.02 \pm 1.28$ \\ 
13.3 & SMA & 215.5 & $35.67 \pm 0.81$ \\ 
13.3 & SMA & 231.5 & $32.94 \pm 1.01$ \\ 
14.36 & SMA & 344.8 & $26.85 \pm 2.22$ \\ 
14.36 & SMA & 360.8 & $26.13 \pm 2.77$ \\ 
14.37 & SMA & 215.5 & $42.05 \pm 0.5$ \\ 
14.37 & SMA & 231.5 & $38.71 \pm 0.58$ \\ 
15.23 & SMA & 225.0 & $30.82 \pm 2.41$ \\ 
15.23 & SMA & 233.0 & $28.64 \pm 4.0$ \\ 
15.23 & SMA & 241.0 & $27.41 \pm 3.21$ \\ 
15.23 & SMA & 249.0 & $15.4 \pm 4.74$ \\ 
17.29 & SMA & 234.6 & $36.57 \pm 1.55$ \\ 
17.29 & SMA & 250.6 & $34.04 \pm 1.81$ \\ 
18.4 & SMA & 217.5 & $52.52 \pm 0.55$ \\ 
18.4 & SMA & 233.5 & $49.32 \pm 0.65$ \\ 
19.25 & SMA & 193.5 & $59.27 \pm 1.49$ \\ 
19.25 & SMA & 202.0 & $56.03 \pm 1.5$ \\ 
19.25 & SMA & 209.5 & $55.09 \pm 1.39$ \\ 
19.25 & SMA & 218.0 & $54.54 \pm 1.33$ \\ 
20.28 & SMA & 215.5 & $50.6 \pm 1.69$ \\ 
20.28 & SMA & 231.5 & $49.16 \pm 1.84$ \\ 
20.28 & SMA & 267.0 & $41.69 \pm 1.62$ \\ 
20.28 & SMA & 283.0 & $37.84 \pm 1.63$ \\ 
24.39 & SMA & 215.5 & $55.57 \pm 0.53$ \\ 
24.39 & SMA & 231.5 & $53.2 \pm 0.6$ \\ 
24.4 & SMA & 333.0 & $23.98 \pm 1.39$ \\ 
24.4 & SMA & 349.0 & $28.46 \pm 1.37$ \\ 
26.26 & SMA & 215.6 & $38.83 \pm 1.2$ \\ 
26.26 & SMA & 231.6 & $34.1 \pm 1.33$ \\ 
31.2 & SMA & 230.6 & $36.76 \pm 1.12$$^a$ \\ 
31.2 & SMA & 246.6 & $31.41 \pm 1.42$$^a$ \\ 
35.34 & SMA & 215.5 & $21.59 \pm 0.89$ \\ 
35.34 & SMA & 231.5 & $20.63 \pm 1.04$ \\ 
36.34 & SMA & 215.5 & $24.32 \pm 1.19$ \\ 
36.34 & SMA & 231.5 & $20.79 \pm 1.42$ \\ 
39.25 & SMA & 217.0 & $18.34 \pm 1.65$ \\ 
39.25 & SMA & 233.0 & $19.74 \pm 1.76$ \\ 
39.26 & SMA & 264.0 & $17.61 \pm 2.79$ \\ 
39.26 & SMA & 280.0 & $8.27 \pm 2.93$ \\ 
41.24 & SMA & 217.0 & $12.58 \pm 1.5$ \\ 
41.24 & SMA & 225.0 & $8.91 \pm 1.9$ \\ 
41.24 & SMA & 233.0 & $15.08 \pm 1.73$ \\ 
41.24 & SMA & 241.0 & $9.64 \pm 2.13$ \\ 
44.24 & SMA & 230.6 & $9.42 \pm 1.61$ \\ 
44.24 & SMA & 234.6 & $8.04 \pm 2.51$ \\ 
44.24 & SMA & 246.3 & $10.43 \pm 2.13$ \\ 
44.24 & SMA & 250.6 & $10.06 \pm 3.24$ \\ 
45.23 & SMA & 217.0 & $8.28 \pm 2.24$ \\ 
45.23 & SMA & 233.0 & $10.55 \pm 2.39$ \\ 
45.23 & SMA & 264.0 & $8.35 \pm 3.27$ \\ 
45.23 & SMA & 280.0 & $5.7 \pm 3.49$ \\ 
47.24 & SMA & 230.6 & $11.47 \pm 2.81$ \\ 
47.24 & SMA & 234.6 & $10.81 \pm 4.39$ \\ 
47.24 & SMA & 246.6 & $11.65 \pm 3.76$ \\ 
47.24 & SMA & 250.6 & $5.6 \pm 5.37$ \\ 
48.31 & SMA & 217.5 & $7.63 \pm 1.11$ \\ 
48.31 & SMA & 233.5 & $5.73 \pm 1.32$ \\ 
76.27 & SMA & 215.5 & $1.33 \pm 0.55$ \\ 
76.27 & SMA & 231.5 & $0.61 \pm 0.63$ \\ 
76.27 & SMA & 335.0 & $-2.27 \pm 1.87$ \\ 
76.27 & SMA & 351.0 & $-0.32 \pm 1.76$ \\ 
10.48 & ATCA & 5.5 & $< 0.15$ \\ 
10.48 & ATCA & 9.0 & $0.27 \pm 0.06$ \\ 
10.48 & ATCA & 34.0 & $5.6 \pm 0.16$ \\ 
13.47 & ATCA & 5.5 & $0.22 \pm 0.05$ \\ 
13.47 & ATCA & 9.0 & $0.52 \pm 0.04$ \\ 
13.47 & ATCA & 16.7 & $1.5 \pm 0.1$ \\ 
13.47 & ATCA & 21.2 & $2.3 \pm 0.3$ \\ 
13.47 & ATCA & 34.0 & $7.6 \pm 0.5$ \\ 
17.47 & ATCA & 5.5 & $0.41 \pm 0.04$ \\ 
17.47 & ATCA & 9.0 & $0.99 \pm 0.03$ \\ 
19.615 & ATCA & 34.0 & $14.26 \pm 0.21$ \\ 
28.44 & ATCA & 34.0 & $30.59 \pm 0.2$ \\ 
34.43 & ATCA & 34.0 & $42.68 \pm 0.19$ \\ 
81.37 & ATCA & 34.0 & $6.97 \pm 0.09$ \\ 
14.03 & ALMA & 336.5 & $29.4 \pm 2.94$ \\ 
14.03 & ALMA & 338.5 & $29.1 \pm 2.91$ \\ 
14.03 & ALMA & 348.5 & $28.49 \pm 2.85$ \\ 
14.03 & ALMA & 350.5 & $28.29 \pm 2.83$ \\ 
14.14 & ALMA & 398.0 & $26.46 \pm 2.65$ \\ 
14.14 & ALMA & 400.0 & $26.21 \pm 2.62$ \\ 
14.14 & ALMA & 410.0 & $25.69 \pm 2.57$ \\ 
14.14 & ALMA & 412.0 & $25.95 \pm 2.6$ \\ 
22.02 & ALMA & 90.5 & $91.18 \pm 4.6$ \\ 
22.02 & ALMA & 92.5 & $92.31 \pm 4.6$ \\ 
22.02 & ALMA & 102.5 & $93.97 \pm 4.7$ \\ 
22.02 & ALMA & 104.5 & $93.57 \pm 4.7$ \\ 
22.04 & ALMA & 138.0 & $85.1 \pm 4.3$ \\ 
22.04 & ALMA & 140.0 & $84.58 \pm 4.2$ \\ 
22.04 & ALMA & 150.0 & $80.62 \pm 4.0$ \\ 
22.04 & ALMA & 152.0 & $79.71 \pm 4.0$ \\ 
23.06 & ALMA & 671.0 & $31.5 \pm 6.3$ \\ 
\enddata 
\tablecomments{${^a}$ Systematic uncertainty 20\% due to uncertain flux calibration}\end{deluxetable} 

\subsection{X-ray observations}
\label{sec:obsxray}

\subsubsection{\swift/XRT}
\label{sec:swift}

The Neil Gehrels Swift Observatory (\swift; \citealt{Gehrels2004}) has been monitoring AT2018cow since June 19,
with both the
Ultraviolet-Optical Telescope (UVOT; \citealt{Roming2005}) and the X-ray Telescope (XRT; \citealt{Burrows2005}).
The transient was well-detected in both instruments (e.g. \citealt{RiveraSandoval2018d}).

We downloaded the \swift/XRT data products (light curves and spectra) using the web-based tools developed by the \swift-XRT team
\citep{Evans2009}.
We used the default values,
but binned the data by observation.
To convert from count rate to flux, we used the absorbed count-to-flux rate set by the spectrum on the same tool, 
$4.26 \times 10^{-11}\,\erg\,\pcmsq\,\pct$.
This assumes a photon index of $\Gamma=1.54$
and a Galactic $N_H$ column of $6.57 \times 10^{20}\,\pcmsq$.

\subsubsection{NuSTAR}
\label{sec:nustar}

The {\em Nuclear Spectroscopic Telescope Array} (\nustar; \citealt{Harrison2013})
comprises two co-aligned telescopes,
Focal Plane Module A (FPMA) and FPMB.
Each is sensitive to X-rays in the 
3--79\,\kev\ range, with slightly different response functions.
\nustar\ observed AT2018cow on four epochs,
and a log of these observations as well as the best-fit spectral model parameters is presented in Table \ref{tab:nustar}.

{\em NuSTAR} data were extracted using \texttt{nustardas\_06Jul17\_v1} from
HEASOFT~6.24. Source photons were extracted from a circle of 60\arcsec\
radius, visually centered on the object. We note that such a large
region, appropriate for {\em NuSTAR} data, includes the transient
as well as the host galaxy. Background photons were
extracted from a non-overlapping circular region with 120\arcsec\ radius
on the same chip. Spectra were grouped to 20 source photons per bin,
ignoring energies below 3~keV and above 80~keV. 

Spectra were analysed in XSPEC (v12.10.0c), using {\em NuSTAR} CALDB files
dated 2018 August 14. \citet{RiveraSandoval2018d} report a low
absorbing column density (N$_\mathrm{H} = 7.0 \times
10^{20}\,\pcmsq$), hence we ignore this component in fitting.
We opt for a simple phenomenological model to describe the spectrum. We
do not fit for a cross-normalisation constant between {\em NuSTAR} FPMA and
FPMB. Epoch 1 (OBSID 90401327002) spectra are not consistent with a
simple power law or a broken power law, hence we fit it with the \texttt{bkn2pow}
model (obtaining spectral breaks at $9.0\pm0.3$\,\kev\ and $11.1\pm0.3\,\kev$).
Spectra of the remaining three epochs are well-fit by a simple,
unabsorbed power law. 

We calculate the flux directly from energies of individual source and
background photons detected, converted into flux using the Ancillary Response Files (ARF) generated by the \nustar\ pipeline. We use a bootstrap method to
estimate the error bars: we draw photons from the data with replacement,
and calculate the source flux from this random sample. By repeating this
process 10000 times for each OBSID and each energy range, we calculate
the 1-sigma error bars on the fluxes. This method gives answers consistent with \texttt{xspec} flux and \texttt{cflux} measurements for bright sources (see
for instance \citealt{Kaspi2014}), but has the advantage of
giving flux measurements without the need to assume a spectral model for
the source. We find that the source is not well-detected in the 40--80\,\kev\ band at the third and fourth epochs.

\begin{deluxetable*}{cc|cc|cccc|cc}
\tabletypesize{\scriptsize}
\tablecaption{\nustar\ flux measurements for AT2018cow, and the spectral model parameters \label{tab:nustar}}
\tablehead{
\colhead{Epoch} & \colhead{OBSID} &
\colhead{Exp. time (ks)} & 
\colhead{$\Delta t$ (d)} &
\multicolumn{4}{c}{Flux ($10^{-12}~\mathrm{erg~cm}^{-2}~\mathrm{s}^{-1}$)} & 
\colhead{Photon Index} & \colhead{$\chi^2$/DOF}
\\
\colhead{} & \colhead{} & \colhead{} & \colhead{} & \colhead{3--10~keV} & \colhead{10--20~keV} & \colhead{20--40~keV} & \colhead{40--80~keV}
}
\startdata
1 & 90401327002\tablenotemark{a} & 32.4 & 7.9 & 4.94 $\pm$ 0.04 & 4.41 $\pm$ 0.10 & 12.21 $\pm$ 0.39 & 21.46 $\pm$ 4.29 & \nodata & 421/443 \\
2 & 90401327004 & 30.0 & 16.8 & 5.21 $\pm$ 0.04 & 4.99 $\pm$ 0.10 &  7.70 $\pm$ 0.33 & 12.80 $\pm$ 4.79 & $1.39 \pm 0.02$ & 424/412 \\
3 & 90401327006 & 31.2 & 28.5 & 1.58 $\pm$ 0.03 & 1.45 $\pm$ 0.06 &  1.74 $\pm$ 0.21 &  \nodata & $1.51 \pm 0.04$ & 174/169 \\
4 & 90401327008 & 33.0 & 36.8 & 1.10 $\pm$ 0.02 & 0.92 $\pm$ 0.05 &  1.02 $\pm$ 0.20 & \nodata & $1.59 \pm 0.05$ & 134/135 \\
\enddata
\tablecomments{Fluxes were measured with a model-independent method.}
\tablenotetext{a}{OBSID 90401327002 is best described by a \texttt{bkn2pow} model with parameters $\Gamma_1 = 1.24\pm 0.05$, $E_1 = 9.0\pm0.3$~keV, $\Gamma_2 = 3.6\pm0.7$, $E_2 = 11.1\pm0.3$~keV, $\Gamma_3 = 0.50 \pm 0.05$. All reported values are for this model.}
\end{deluxetable*}

\section{Basic properties of the shock}
\label{sec:modeling}

\subsection{Light curve}
\label{sec:lightcurve}

\begin{figure}[hp]
\centering
\includegraphics[scale=0.75]{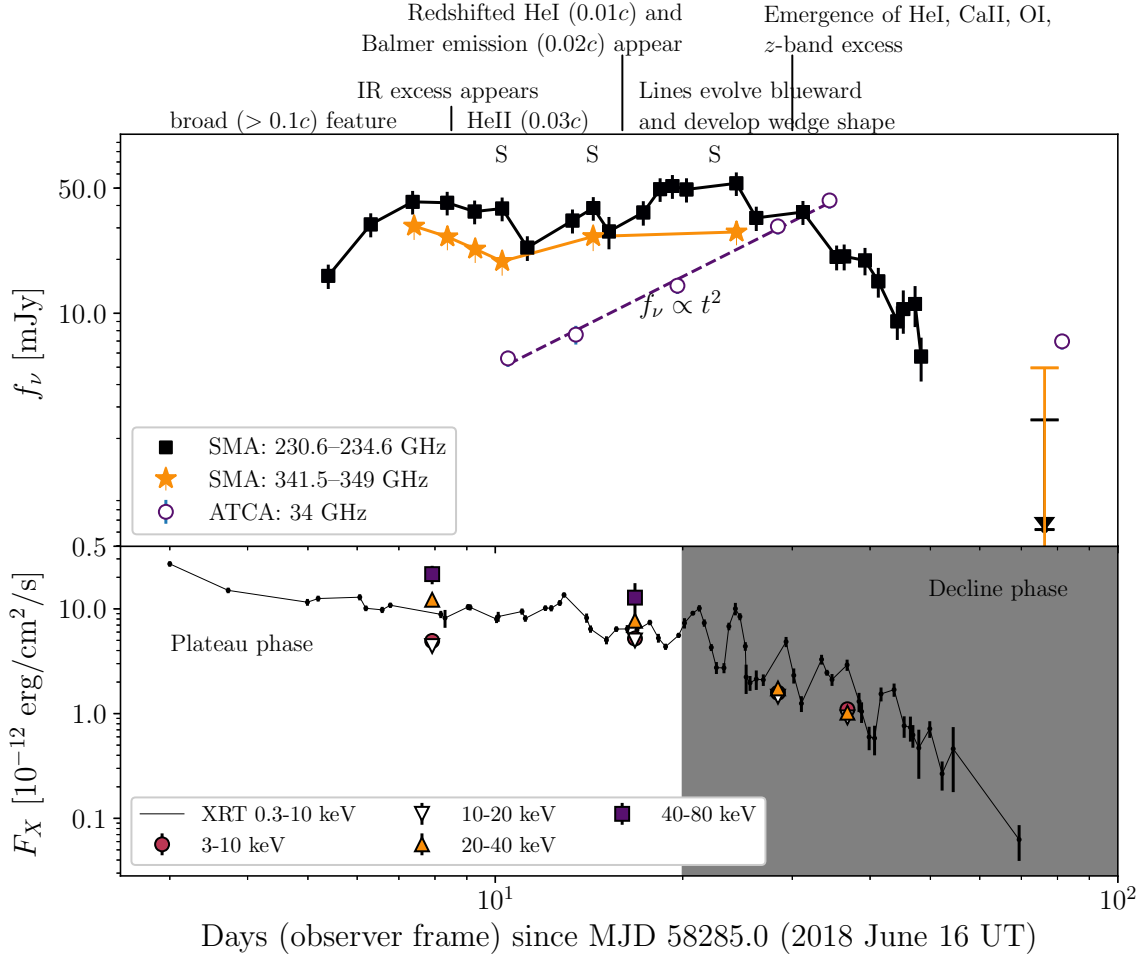}
\caption{
(Top panel) Submillimeter (SMA) through radio (ATCA) light curves of AT2018cow, with a timeline of the evolution of the UVOIR spectra (based on \citealt{Perley2018}) shown above. There were four SMA observations with no frequency tunings in the ranges shown. For these, we took the closest value to 231.5\,\ghz\ (243.3\,\ghz\ for Days 9, 10, and 11; 218\,\ghz\ for Day 19) and scaled them to 231.5\,\ghz\ assuming a spectral index $F_\nu \propto \nu^{-1}$.
We scaled all SMA fluxes so that the reference quasar 1635+381 would have the value of its mean flux at that frequency.
The uncertainties shown on the SMA data represent a combination of formal uncertainties and 15\% systematic uncertainties, which is a conservative estimate.
Non-detections are represented as a 3-$\sigma$ upper limit (horizontal bar) and a vertical arrow down to the measurement.
The upper limit measurement at 350.1\,\ghz\ is $-0.32$, below the limit of the panel.
The error bars shown on the ATCA data are a combination of formal uncertainties and an estimated 10\% systematic uncertainty.
The ATCA 34\,\ghz\ measurements rise as $t^{2}$,
shown as a dotted line.
The full set of SMA light curves for all frequency tunings are shown in Appendix \ref{sec:appendix-smalc}.
The letters `S' on the top demarcate the epochs with spectra shown in Figure~\ref{fig:spec}.
(Bottom panel) X-ray light curve from \swift/XRT together with four epochs of \nustar\ observations.
The last two \nustar\ epochs have a non-detection in the highest-frequency band (40--80\,keV).
We denote two distinct phases of the X-ray light curve, the plateau phase and the decline phase, discussed in detail in Section \ref{sec:modeling}.}
\label{fig:lc}
\end{figure}

The radio and X-ray light curves are shown in Figure~\ref{fig:lc}.
The 230\,\ghz\ light curve rises (the first observation of a millimeter transient in its rise phase)
and then shows significant variability,
presumably from inhomogeneities in the surrounding medium.
We have tentative evidence that the rise is at least in part due to a decreasing peak frequency: at $\Delta t=5$--6\,\days, the flux is marginally higher at 231.5\,\ghz\ than at 215.5\,\ghz, and at $\Delta t=7$--8\,\days,
it seems that the peak may have been within the SMA observing bands. However, the position of the peak is ill-constrained;
future early observations would benefit from observations at more frequencies.

By $\Delta t=50\,\days$,
the radio flux has diminished both
due to the peak frequency shifting to lower frequencies,
{\it and} to a decay in the peak flux.
Specifically, 
the peak of the 15\,\ghz\ light curve is 19\,\mjy\ around 47 days  (A. Horesh, personal communication), substantially less luminous than the peak of the 230\,\ghz\ or the 34\,\ghz\ light curve.
As we discuss in Section \ref{sec:luminosity}, this diminishing peak flux suggests that the interaction itself is diminishing, and enables us to constrain the size of the ``circum-bubble'' of material.

The X-ray light curve seems to have two distinct phases.
We call the first phase ($\Delta t \lesssim 20\,$days)
the {\it plateau} phase because the X-ray emission is relatively flat.
The second phase, which we call the \emph{decline} phase,
begins around $\Delta t \approx 20\,$days.
During this period, the X-ray emission exhibits an overall steep decline, but also exhibits strong variation (by factors of up to 10) on shorter timescales (see also \citealt{RiveraSandoval2018d,Kuin2018,Perley2018}).

\begin{figure}[ht]
\centering
\includegraphics[scale=0.8]{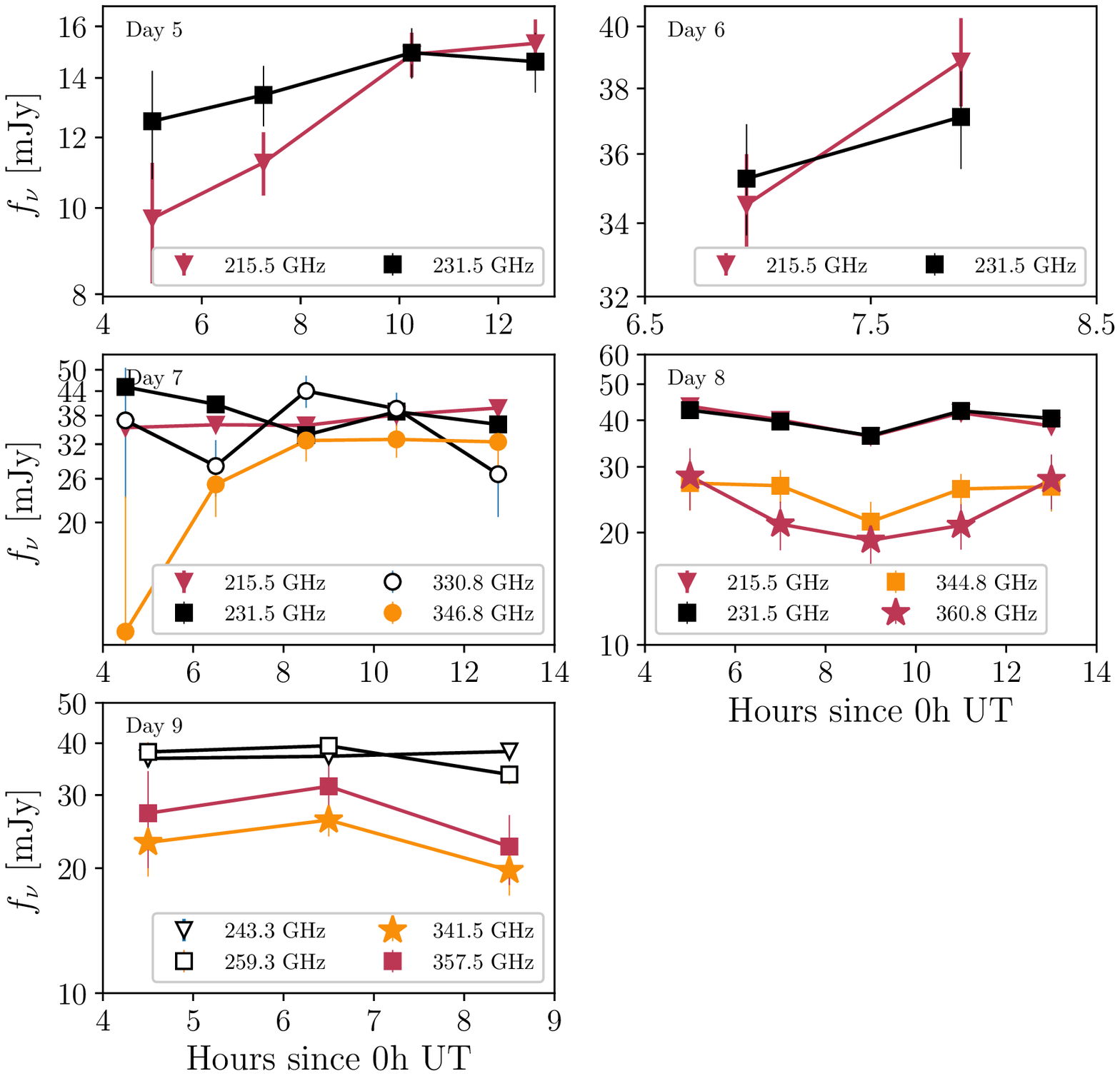}
\caption{Zoomed-in light curves for the first five days of SMA observations. These were the only tracks long enough for binning in time.
}
\label{fig:lc-zoom}
\end{figure}

We use the shortest timescale of variability in the 230\,\ghz\ light curve
to infer the size of the radio-emitting region,
and do the same for the X-ray emission in Section \ref{sec:xray}.
On Days 5--6, the 230\,\ghz\ flux changed by order unity in one day,
setting a length scale for the source size of
$\Delta R = c \Delta t = 2.6 \times 10^{15}\,\cm$ (170\,AU).
We find no evidence for
shorter-timescale variability in our long SMA tracks from the first few days of observations (Figure~\ref{fig:lc-zoom}).

Together with the 230\,\ghz\ flux density ($S_\nu \approx 30\,\mjy$) and the distance ($d=60\,\mpc$) we infer an angular size of $\theta =2.8\,\mas$
and a brightness temperature of

\begin{equation}
    T_B = \frac{S_\nu c^2}{2 k \nu^2 \Delta \Omega} \gtrsim 3 \times 10^{10}\,\kelvin
\end{equation}

\noindent where $\Delta\Omega=\pi\theta^2$. This brightness temperature is close to the typical rest-frame equipartition brightness temperatures of the most compact radio sources, $T_B\sim5\times10^{10}$\,K \citep{Readhead1994}.

\subsection{Modeling the radio to sub-millimeter SED}
\label{sec:sed}

\begin{figure}[ht]
\centering
\includegraphics[scale=0.83]{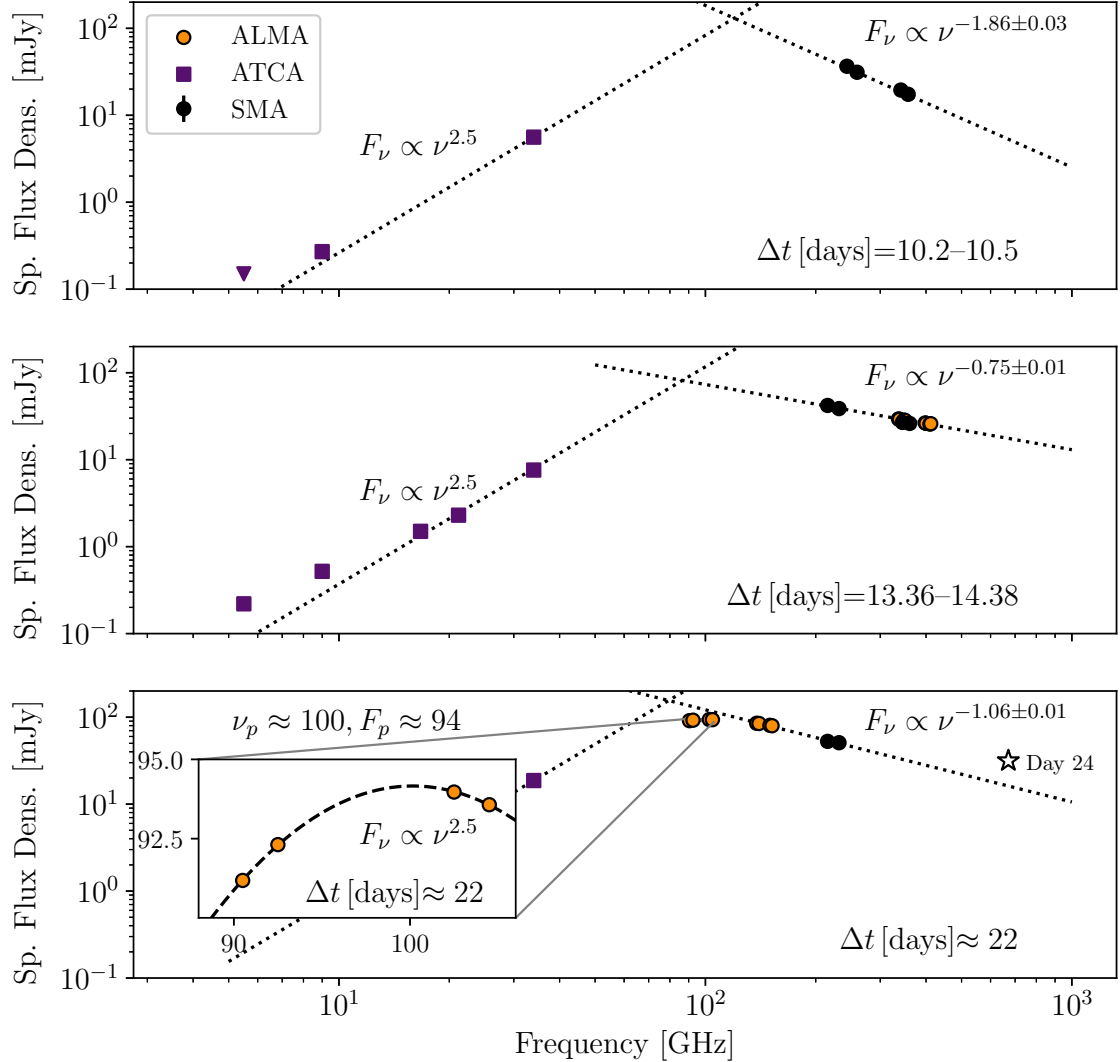}
\caption{Spectrum of AT2018cow at three epochs. 
In the top panel, we plot the Day 10 data as presented in Table \ref{tab:flux}.
In the middle panel,
we plot the ATCA data from Day 13
and the SMA and ALMA data from Day 14.
In the bottom panel,
we plot the ALMA data from Day 22, interpolate the SMA data between Day 20 and Day 24 at 215.5\,\ghz\ and 231.5\,\ghz,
and interpolate the ATCA data at 34\,\ghz\
(since it varies smoothly; Figure \ref{fig:lc}).
We also show the Band 9 measurement from Day 24 as a star.
The ATCA data is consistent with a self-absorbed spectral index ($F_\nu \propto \nu^{5/2}$) with an excess at lower frequencies.
The peak frequency is resolved on Day 22 with ALMA observations at Band 3 (see inset). 
To measure the optically thin spectral index, we performed a least squares fit in log space. 
To estimate the uncertainty on the spectral index, we performed a Monte Carlo analysis, sampling $10^4$ times to measure the standard deviation of the resulting spectral index.
On Day 10, we used an uncertainty of 15\% for each SMA measurement.
On Days 14 and 22, we used 10\% uncertainty for each ALMA measurement and 20\% for each SMA measurement (to take into account the much longer length of the SMA tracks).
Uncertainties are too small to be visible on this plot, except for the inset panel, where we do not display them.}
\label{fig:spec}
\end{figure}

The shape of the radio to sub-millimeter SED (Figure \ref{fig:spec}),
together with the high brightness temperature implied by the luminosity and variability timescale (Section \ref{sec:lightcurve}),
can only be explained by non-thermal emission \citep{Readhead1994}.
The observed spectrum is assumed to arise from a 
population of electrons with a power-law number distribution in Lorentz factor $\gamma_e$, with some minimum Lorentz factor $\gamma_m$ and electron energy power index $p$:
\begin{equation}
\label{eq:powerlaw}
    \frac{dN(\gamma_e)}{d\gamma_e} \propto \gamma_e^{-p}, \qquad \gamma_e \geq \gamma_m.
\end{equation}

\noindent 
As argued below, we expect an adiabatic strong shock moving into a weakly magnetized, ionized medium at a non-relativistic speed. First-order Fermi acceleration gives $p=(r+2)/(r-1)$, where $r$ is the compression ratio of the shock.  A strong matter-dominated shock has $r=4$, hence $p=2$ \citep{Blandford1987}.  However the back-reaction of the accelerated particles decelerates the gas flow, weakening the gas dynamic subshock and reducing the compression ratio from the strong shock $r=3$, so typical $2.5 < p < 3$ are obtained in both simulations and astrophysical data \citep{Jones1991}.  Quasi-perpendicular magnetized and relativistic shocks are more subtle, since some particles cannot return along field lines after their first shock crossing, but the limiting value is $p \sim 2.3$ \citep{Pelletier2017}.

Equation \ref{eq:gammam} provides an expression for $\gamma_m$.
Behind the shock (velocity $v$) some fraction $\epsilon_e$ of the total energy density goes into accelerating electrons. Conserving shock energy flux (using the low-velocity approximation $\gamma \approx 1 + \frac{1}{2}\beta^2$) gives,

\begin{equation}
\label{eq:gammam}
    \gamma_m - 1 \approx \epsilon_e \frac{m_p}{m_e} \frac{v^2}{c^2}.
\end{equation}

The value of $\gamma_m$ is large for relativistic shocks, e.g. in GRBs. But we will see that for this source ($v/c\sim 0.1$, $\epsilon_e\sim 0.1$), the bulk of the electrons are just mildly relativistic ($\gamma_m\sim 2-3$). For ordinary supernova shocks $\gamma_m$ is always non-relativistic ($\gamma_m-1<1$).  Thus in the parameter estimations below, we follow supernova convention and assume that the relativistic electrons follow a power-law distribution down to a fixed 
$\gamma_m$ \citep{Chevalier1982,Chevalier1998,Kulkarni1998,Frail2000,Soderberg2005}. We apply $\epsilon_e$ {\em only} to this relativistic power-law, not to the nonrelativistic thermal distribution of shock-heated particles at lower energy.

We now describe each of the break frequencies that
characterize the observed  spectrum.
First, the characteristic synchrotron frequency $\nu_m$ emitted by the minimum energy electrons is:

\begin{equation}
\label{eq:num}
    \nu_m = \gamma_m^2 \nu_g
\end{equation}

\noindent where $\nu_g$ is the gyrofrequency,

\begin{equation}
\label{eq:nug}
    \nu_g = \frac{q_e B}{2 \pi m_e c}
\end{equation}

\noindent and $q_e$ is the unit charge, $B$ is the magnetic field strength, $m_e$ is the electron mass, and $c$ is the speed of light.

Next, there is the cooling frequency $\nu_c \equiv \nu(\gamma_c)$,
the frequency below which electrons have lost the equivalent of their total energies to radiation via cooling.
In general, the timescale for synchrotron cooling depends on the Lorentz factor as $t \propto \gamma_e^{-1}$. 
Thus, electrons radiating at higher frequencies cool more quickly. Separately, 
electrons could also lose energy by Compton upscattering of ambient (low energy) photons -- the so-called Inverse Compton (IC) scattering.
In Section \ref{sec:xray},
we find that IC scattering dominates at early times
and that synchrotron losses dominate at later times,
and that the transition is at $t\approx13\,\days$.

At $\Delta t>13\,\days$, electrons with $\gamma_e > \gamma_c$ cool principally by synchrotron radiation to $\gamma_c$ in a time $t$, where

\begin{equation}
    \gamma_c = \frac{6 \pi m_e c}{\sigma_T B^2 t}.
    \label{eq:gammac}
\end{equation}

For $t<13\,\days$, Compton cooling on the UVOIR flux exceeds the synchrotron cooling rate by a factor $\sim (t/10\,\days)^{-5/2}$, and $\gamma_c$ is correspondingly lowered. The cooled electrons emit around the characteristic synchrotron frequency

\begin{equation}
\nu_c = \gamma_c^2 \nu_g.
\label{eq:nuc}
\end{equation}

Next, the self-absorption frequency $\nu_a$ is the frequency at which the optical depth to synchrotron self-absorption is unity. 
The rise at 34\,\ghz\ obeys a $f_\nu \propto t^2$ power law (as shown in Figure \ref{fig:lc}), consistent with the optically thick spectral index we measure (Figure \ref{fig:spec}).
This indicates that the self-absorption frequency is above the ATCA bands ($\nu_a > 34\,\ghz$).
Figure \ref{fig:spec} also shows that
the emission in the SMA bands is optically thin at $\Delta t \gtrsim 10\,\days$,
constraining the self-absorption frequency to be $\nu_a < 230\,\ghz$.

On Day 22, we resolve the peak of the SED with our ALMA data. We denote the peak frequency $\nu_p$ and the flux at the peak frequency $F_p$, and find $\nu_p \approx \nupeak$ and $F_p \approx \fpeak$.
Motivated by the observation of optically thick emission at $\nu<\nu_p$, we assume that $\nu_p = \nu_a$,
and adopt the framework in \citet{Chevalier1998} (hereafter referred to as C98) to estimate properties of the shock at this epoch.
These properties are summarized in Table \ref{tab:measurements},
and outlined in detail below.

Following equation~(11) and equation~(12) in C98, the outer shock radius $R_p$ can be estimated as

\begin{equation}
R_p = \left[ \frac{6 c_6^{p+5} F_p^{p+6} D^{2p+12}}{(\epsilon_e/\epsilon_B) f(p-2) \pi^{p+5} c_5^{p+6} E_l^{p-2}} \right]^{1/(2p+13) }
\left( \frac{\nu_p}{2 c_1} \right)^{-1}
\label{eq:chev-R}
\end{equation}

\noindent and the magnetic field can be estimated as

\begin{equation}
B_p = \left[ \frac{36 \pi^3 c_5}{(\epsilon_e/\epsilon_B)^2 f^2 (p-2)^2 c_6^3 E_l^{2(p-2)} F_p D^2 } \right]^{2/(2p+13) }
\left( \frac{\nu_p}{2 c_1} \right)
\label{eq:chev-B}
\end{equation}

\noindent where, as in Equation \ref{eq:powerlaw},
$p$ is the electron energy index.
Note that C98 use $\gamma$ for the electron energy power index.
We use $p$ instead and $\gamma$ for the Lorentz factor.
The constant $c_1=6.27 \times 10^{18}$ in cgs units, and the constants $c_5$ and $c_6$ are tabulated as a function of $p$ on page 232 of \citet{Pacholczyk1970}.
$D$ is the distance to the source,
$E_l=0.51\,$MeV is the electron rest mass energy,
and $\epsilon_e/\epsilon_B$ is the ratio of energy density in electrons to energy density in magnetic fields (in C98 this ratio is parameterized as $\alpha$, but we use $\alpha$ as the optically thin spectral index of the radio SED.) 
Finally, $f$ is the filling factor:
the emitting region is approximated as a planar region with thickness $s$ and area in the sky $\pi R^2$,
and thus a volume $\pi R^2 s$,
which can be characterized as a spherical emitting volume 
$V = 4 \pi f R^3/3 = \pi R^2 s$.

On Day 22, we measure $\alpha=-1.1$ where $F_\nu \propto \nu^{\alpha}$,
which corresponds to $p=3.2$.
Later in this section we show that our sub-millimeter observations lie above the cooling frequency, and therefore that the index of the \emph{source} function of electrons is $p_s=2.2$. However the C98 prescription considers a distribution as it exists when the electrons are observed,
from a combination of the initial acceleration and the energy losses (to cooling).
So, we proceed with $p=3.2$, and discuss this unusual regime in Section \ref{sec:breakfreq}.
The closest value of $p$ in the table in \citet{Pacholczyk1970} is $p=3$, so we use this value to select the constants (and note that, as stated in C98, the results do not depend strongly on the value of $p$.)
With this, equation~\ref{eq:chev-R} and equation~\ref{eq:chev-B} reduce to equation~(13) and equation~(14) in C98, respectively, reproduced here:

\begin{equation}
\label{eq:chev-R-short}
    R_p = 8.8 \times 10^{15}\,
    \left(\frac{\epsilon_e}{\epsilon_B}\right)^{-1/19} 
    \left( \frac{f}{0.5} \right)^{-1/19}
    \left( \frac{F_p}{\jy} \right)^{9/19}
    \left( \frac{D}{\mpc} \right)^{18/19}
    \left( \frac{\nu_p}{5\,\ghz} \right)^{-1}\,\cm,
\end{equation}

\begin{equation}
\label{eq:chev-B-short}
    B_p = 0.58\,
    \left(\frac{\epsilon_e}{\epsilon_B}\right)^{-4/19} 
    \left( \frac{f}{0.5} \right)^{-4/19}
    \left( \frac{F_p}{\jy} \right)^{-2/19}
    \left( \frac{D}{\mpc} \right)^{-4/19}
    \left( \frac{\nu_p}{5\,\ghz} \right)\,\gauss.
\end{equation}

Next we estimate the total energy $U$.
For $p=3$,
equation~\ref{eq:chev-R-short} and equation~\ref{eq:chev-B-short}
can be combined into the following expression for $U=U_B/\epsilon_B$,
\begin{equation}
\label{eq:energy}
\begin{split}
    U
    &=\frac{1}{\epsilon_B}\frac{4\pi}{3} fR^3 \left( \frac{B^2}{8\pi} \right) \\
    &= (1.9 \times 10^{46}\,\erg) 
    \,\frac{1}{\epsilon_B}\left(\frac{\epsilon_e}{\epsilon_B}\right)^{-11/19}
    \left(\frac{f}{0.5}\right)^{8/19}
    \left(\frac{F_p}{\jy}\right)^{23/19}
    \left(\frac{D}{\mpc}\right)^{46/19}
    \left(\frac{\nu_p}{5\,\ghz}\right)^{-1}.
\end{split}
\end{equation}

Following C98 we take $f=0.5$, but the dependence on this parameter is weak.
In choosing $\epsilon_B$ and $\epsilon_e$
there are several normalizations (or assumptions)
used in the literature. As a result the inferred energy
can vary enormously (see Section \ref{sec:velocity-energy} for further details).
For now, we follow \citet{Soderberg2010}
in setting $\epsilon_e=\epsilon_B=1/3$
(in other words, that energy is equally partitioned between electrons, protons, and magnetic fields).
With all of these choices,
we find that at $\Delta t \approx 22\,\days$,
$R_p \approx \radius$
and $B_p \approx \bfield$.
We find that the total energy $U \approx \energy$.
Assuming 10\% uncertainties in $F_{p}$ and $\nu_{p}$
and a 50\% uncertainty in $p$,
a Monte Carlo with 10,000 samples 
gives uncertainties of 0.15--0.3\,dex in these derived parameters.
Our results are robust to departures from equipartition given the large penalty in the required energy \citep{Readhead1994}.

The mean velocity up to $\Delta t \approx 22\,\days$ is $v=R_p/t_p=\velocity$.
We can write a general expression for $v/c$
(taking $L_p = 4 \pi F_p D^2$, noting that $4\pi\mbox{Jy Mpc}^2=1.2\times 10^{27}\mbox{erg s}^{-1}\mbox{Hz}^{-1}$):

\begin{equation}
    v/c \approx \left(\frac{\epsilon_e}{\epsilon_B}\right)^{-1/19} 
    \left( \frac{f}{0.5} \right)^{-1/19}
    \left( \frac{L_p}{10^{26}\,\erg\,\psec\,\phz} \right)^{9/19}
    \left( \frac{\nu_p}{5\,\ghz} \right)^{-1}
    \left( \frac{t_p}{1\,\days} \right)^{-1}.
\end{equation}

Furthermore, from the $t^2$ rise at 34\,\ghz\ (Figure \ref{fig:lc})
we can infer that the radius increases as $R \propto t$
and therefore that the velocity $v = dR/dt$ is constant.
We put this derived energy and velocity into the context of other energetic transients in Section \ref{sec:velocity-energy}.

Next, we estimate the density of the medium into which the forward shock is propagating.
The ejecta expands into the medium with velocity $v_1$, producing a shock front (a discontinuity in pressure, density, and temperature) with shock-heated ejecta immediately behind this front.
Conservation of momentum across this (forward) shock front requires that 

\begin{equation}
     P_\mathrm{1} + \rho_\mathrm{1} v_\mathrm{1}^2 = P_2 + \rho_2 v_2^2
\end{equation}

\noindent where $P$ is pressure (not to be confused with $p$ used as the power-law index for the electron energy distribution). The subscript 1 refers to the upstream medium (the ambient CSM) and the subscript 2 refers to the downstream medium (the shocked ejecta).
Far upstream, the pressure can be taken to be 0,
and in the limit of strong shocks (for a monatomic gas) $\rho_2/\rho_1=v_1/v_2=4$.
Thus this can be simplified to

\begin{equation}
\label{eq:density}
    \frac{3\rho_1 v_\mathrm{1}^2}{4} = P_2.
\end{equation}

\noindent If the medium is composed of fully ionized hydrogen, $\mu_p=1$ and the number densities of protons and electrons are equal ($n_p=n_e$). 
Using equation~\ref{eq:density} together with equation~\ref{eq:chev-B-short}, as well as the relations $P_2=(1/\epsilon_B)B^2/8\pi$ and $\rho_1=\mu_p m_p n_e$,

\begin{equation}
\begin{split}
    n_e \approx (20\,\pcmcub)\,&
    \left(\frac{1}{\epsilon_B} \right)\left(\frac{\epsilon_e}{\epsilon_B}\right)^{-6/19} 
    \left( \frac{f}{0.5} \right)^{-6/19} 
    \left( \frac{L_p}{10^{26}\,\erg\,\psec\,\phz} \right)^{-22/19}\\
    & \times \left( \frac{\nu_p}{5\,\ghz} \right)^{4}
    \left( \frac{t_p}{1\,\days} \right)^{2}.
\end{split}
\end{equation}

\begin{deluxetable}{lrr}
\tablecaption{Quantities derived from Day 22 measurements, using different equipartition assumptions. In the text unless otherwise stated we use $\epsilon_e=\epsilon_B=1/3$
\label{tab:measurements}}
\tablehead{\colhead{Parameter} & \colhead{$\epsilon_e=\epsilon_B=1/3$} & \colhead{$\epsilon_e=0.1, \epsilon_B=0.01$} } 
\tabletypesize{\normalsize} 
\startdata
$\nu_a = \nu_p$ (GHz) & 100 & 100 \\
$F_{\nu,p}$ (mJy) & 94 & 94 \\
$r$ ($10^{15}$\,cm) & 7 & 6 \\
$v/c$ & 0.13 & 0.11 \\
$B$ (G) & 6 & 4 \\
$U(10^{48}\,$erg) & $4$ & 35 \\
$n_e$ ($10^5\,$cm$^{-3}$) & $3$ & 41 \\
$\nu_c$ (GHz) & $2$ & 8 \\
\enddata
\end{deluxetable}

We find that the number density of electrons at $\Delta t\approx22\,\days$ is
$n_e \approx \density$.
We note that the strong jump conditions
used here assume $\gamma=5/3$, and that there is a correction for the contribution of a relativistic ($\gamma=4/3$) component. \citet{Chevalier1983}
quantify this correction using the factor $w$, the ratio of the relativistic pressure to the total pressure. In the most extreme case ($w=1$) the correction is small, only a factor of 1.14 in $n_e$. This is negligible compared to our uncertainties.

At such a high density, the optical depth to free-free absorption $\tau_{\mathrm{ff}}$ might be expected to have a significant effect on the shape of the spectrum at low radio frequencies \citep{Lundqvist1988}.
From \citet{Lang1999},
we have 

\begin{equation}
    \tau_\mathrm{ff} = 8.235 \times 10^{-2}
    \left(\frac{T_e}{\kelvin}\right)^{-1.35}
    \left(\frac{\nu}{\ghz}\right)^{-2.1}
    \int \left(\frac{N_e}{\pcmcub}\right)^2 \left(\frac{dl}{\pc}\right)
\end{equation}

\noindent which, with our measured values of $n_e$ and $R$ on Day 22,
gives the characteristic value

\begin{equation}
    \tilde{\tau}_{\mathrm{ff}} = 
    68
    \left(\frac{T_e}{8000 \kelvin}\right)^{-1.35}
    \left(\frac{\nu}{\ghz}\right)^{-2.1}.
    \label{eq:freefreeoptd}
\end{equation}

However, in AT2018cow the gas through which the shock is propagating is {\em not}
at normal HII-region temperatures of $\sim 10^4\mbox{K}$.
The UV and X-ray photons emitted at early times will completely ionize and
Compton heat any surrounding gas: for gas at the density and radius given in Table~\ref{tab:measurements}, the lifetime to photoionization of a neutral hydrogen atom is less than 0.01\,\second, while the recombination time is years\footnote{
For much lower temperatures $T \sim 10^4 \,\kelvin$,
the Case B (high-density limit) recombination coefficient is
$\alpha_B(T=10^4\,\kelvin) = 2.6 \times 10^{-13}\,\cmcub\,\psec$ \citep{Draine2011},
and the timescale is $t_\mathrm{recomb} = 1/(\alpha_B n_e)$.
For $n_e = 3\times10^{5}\,\pcmcub$, $t_\mathrm{recomb} \approx 250\,\days$.
This timescale becomes even longer for the expected higher temperatures.
}. 
Compton heating of the electrons increases their temperature at the rate

\begin{equation}
    \frac{d (3/2)kT_e}{dt}= H = \frac{\sigma_T}{m_e c^2} \int_0^\infty\frac{h\nu L_\nu f_{KN}(h\nu/m_e c^2)}{4\pi R^2}d\nu \,
\end{equation}

\noindent where the Klein-Nishina correction $f_{KN}(x)\simeq 1-21x/5+O(x^2)$.\footnote{Expressions for $f_{KN}$ for cold electrons are given e.g., in equation~(A1) of \citet{Sazonov2004} and equation~(5) of \citet{Madau1999}, and
for finite temperature electrons in equation~(14) of \citet{Guilbert1986}.}
Even though the blackbody ($T=30,000\,\kelvin$) luminosity at $\Delta t=3\,\days$ is 100 times larger than the coeval X-ray luminosity \citep{Perley2018}, the Compton heating is dominated by the 10--100$\,\kev$ X-ray flux, and
we find, for $3<\Delta t<20\,\days$, gas at the density and radius given in Table~\ref{tab:measurements} has

\begin{equation}
    T_e(t)\simeq 1.0\times 10^6\,\kelvin(t/3\,\days)^{0.6}
\end{equation}

Given the spectral evolution shown in \citet{Perley2018}, the Compton temperature (at which Compton heating balances Compton cooling) is $T_c\sim 2.5\times 10^6\,\kelvin$ on day 3, hardening to
$T_c\sim 1.8\times 10^7\,\kelvin$ on day 20 since the blackbody UV flux drops as $t^{-2.5}$, while the hard X-ray flux drops much more slowly.
At these high Compton-heated temperatures $T_e\sim 10^6\,\kelvin$, the free-free absorption optical depth given by equation~\ref{eq:freefreeoptd} only rises above unity
below frequencies of 300\,MHz,
accessible to facilities like LOFAR.

Next, we estimate the luminosity from free-free emission
of the ionized gas \citep{Lang1999}:

\begin{equation}
    L \approx 1.43 \times 10^{-27} \, 
    n_e n_i T^{1/2} V Z^2 g \, \lum
\end{equation}

\noindent where $n_i$ is the number density of ions, $Z$ is the atomic number, and $g\approx1$ is the Gaunt factor, a quantum mechanical correction.
Assuming that the gas is completely ionized out to the light travel sphere at 22 days ($R=6 \times 10^{16}\,\cm$), we have $n_e = n_i$ in the region of interest. We also take $Z=1$.
With the inferred density (Table \ref{tab:measurements})
we find $L \approx 9 \times 10^{37}\,\lum$,
so the contribution to the observed X-ray luminosity is negligible.

Finally, we estimate the different break frequencies,
beginning with $\nu_m$.
Using equation~\ref{eq:gammam},
taking $\epsilon_e = 1/3$ and using our inferred $\beta=v/c$ from Day 22, $\gamma_m \approx 5$.
Next, using equation~\ref{eq:nug} and our measured value of $B$, $\nu_g \approx 17\,\mhz$.
Equation~\ref{eq:num} thus gives $\nu_m \approx 0.4\,\ghz$, substantially below our peak frequency.
The spectral index at $\nu_m < \nu < \nu_a$ is $\nu^{5/2}$ \citep{RL},
which we show as dotted lines in Figure \ref{fig:spec}.
Clearly, the lowest frequency fluxes are in 
excess of $\nu^{5/2}$ extrapolation.
This naturally occurs if the source is inhomogeneous (e.g. magnetic field and/or particle energy density decreasing outwards). It can also arise even for a perfectly homogenous source because the energy spectrum of the radiating electrons is not a pure power-law: note that $\nu_a>\nu_c>\nu_m$, so even beyond the Maxwellian-like peak at $\gamma_m$, the spectrum is convex, steepening with energy above $\gamma_c$. 
These both produce self-absorbed spectra flatter than $\nu^{5/2}$ (see Section 6.8 in \citet{RL} and \citet{deKool1989} for model calculations).

The cooling frequency due to synchrotron radiation is determined by
equation~\ref{eq:gammac} and equation~\ref{eq:nuc}.
We find $\gamma_c\approx11$, giving a cooling frequency
$\nu_c\approx\nucool$.
The relative contributions to electron cooling
from synchrotron radiation and IC scattering
are determined by
the ratio between the radiation energy
density and the magnetic energy density.
On Day 22, the bolometric luminosity as measured in the UVOIR is $5 \times 10^{42}\,\lum$ \citep{Perley2018},
so the radiation density is $u_{\mathrm{ph}} = 0.26\,\erg\,\pcmcub$.
The magnetic energy density on the same day is
$u_B = B^2/8\pi\approx 1.5\,\erg\,\pcmcub.$
Thus synchrotron radiation is the dominant
cooling mechanism, with a roughly 10\% contribution
from IC scattering.

At this epoch, the cooling timescale
$t_{\mathrm{cool}}=(\gamma_e m_e c^2)/(\frac{4}{3}\sigma_T u_B \gamma_e^2 c)=240\,\days/\gamma_e$
for an electron with $\gamma_e$,
which is roughly 80 for an electron radiating at 100\,\ghz.
So on Day 22 the cooling timescale
is shorter than the timescale on which we are
observing the source.
This means that continuous re-acceleration
of the electrons is required,
which could be provided by ongoing shock interaction.

As stated in Section \ref{sec:lightcurve}, it seems that $\nu_p$ during the rise phase ($\Delta t\approx5$--8\,\days) was above or within the SMA observing bands.
Using the peak observed flux and frequency as a lower limit on the peak flux and peak frequency, respectively, we consistently find that $v\approx0.1c$, albeit with a decreasing $n_e$ ($3\times10^{6}\,\pcmcub$ at $\Delta t\approx5\,\days$).

\section{Implications of shock properties}
\label{sec:implications}

\subsection{AT2018cow in velocity-energy space, and a discussion of epsilons}
\label{sec:velocity-energy}

It is challenging to directly compare the energy of AT2018cow
to that of other classes of radio-luminous transients,
because there are several
conventions that produce discrepant results.
In particular, the energy partition fractions $\epsilon_B$ and $\epsilon_e$ are important for determining the total amount of energy in the shock, but are difficult to measure.

In the classical gamma-ray burst (GRB) literature,
$\epsilon_e$ has been consistently measured to be $\epsilon_e \approx 0.2$, within a factor of 2,
while values of $\epsilon_B$ have much wider spreads, with a median value of $3 \times 10^{-5}$ but a distribution spanning four orders of magnitude \citep{Kumar2015}.
\citet{Kumar2010} constrain 
$\epsilon_B \sim 10^{-6}$ for a CSM density of $0.1\,\pcmcub$,
and an even smaller value for higher densities.
One of the best-observed GRB afterglows is GRB\,130427A,
and from modeling the evolving spectrum
\citet{Perley2014} find $0.03 < \epsilon_B < 1/3$ and $0.14 < \epsilon_e < 1/3$.

There are different approaches to modeling
for the handful of low-luminosity GRBs (LLGRBs)
discovered to date.
For LLGRB\,980425/SN\,1998bw, \citet{Kulkarni1998}
invoke equipartition ($\epsilon_B = \epsilon_e = 0.5$). 
For LLGRB\,031203/SN2003\,lw, \citet{Soderberg2004} use models from \citet{Sari1998} and \citet{Granot2002} together with the cooling frequency inferred from X-ray observations to estimate $\epsilon_e = 0.4$ and $\epsilon_B = 0.2$.
For LLGRB\,060218/SN\,2006aj, \citet{Soderberg2006} use the same prescription as was used in SN\,1998bw.
Finally, for LLGRB\,100316D/SN\,2010bh, \citet{Margutti2013} set $\epsilon_B=0.01$ and allow $\epsilon_e$ to vary from 0.01--0.1. 

For Type II and Type Ibc radio supernovae,
approaches range from using the SN\,1998bw convention
(i.e. $\epsilon_B=\epsilon_e=0.5$; \citealt{Soderberg2005,Horesh2013})
to $\epsilon_B=\epsilon_e=0.1$ (e.g. \citealt{Chevalier2006,Soderberg2006,Salas2013})
to $\epsilon_B=\epsilon_e=1/3$ for the relativistic supernova SN\,2009bb \citep{Soderberg2010}.

In this work, we follow the convention in \citet{Soderberg2010}
so that we can compare our velocity-energy diagram
to the corresponding diagram (Figure 4) in that paper.
To put all transients on the same scale,
we take the peak frequency, peak luminosity, and peak time for each event,
and run them through the same equations that we used to infer the shock properties of AT2018cow.
Note that we do not vary the values of $c_5, c_6$, $p$, but these are all very small corrections, whereas the effect of $\epsilon_B$ and the ratio $\epsilon_e/\epsilon_B$ in estimating the energy is large.
The details of how we selected the peak values for each event are in Appendix \ref{sec:details}.
When possible, we use the peak of the SED at a particular epoch. However, for most events, we use the peak flux density corresponding to a certain frequency, because well-sampled SEDs are rare.

\begin{figure}[ht]
\centering
\includegraphics[scale=0.8]{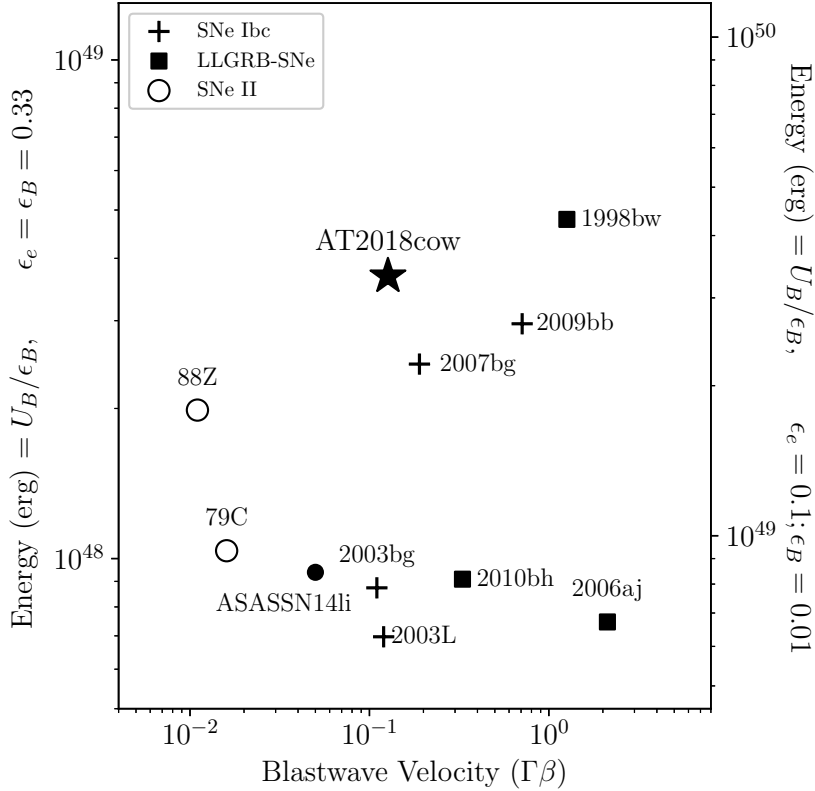}
\caption{AT2018cow in velocity-energy space, compared to other classes of radio-luminous transients: TDEs (filled circles), Ibc supernovae (crosses), SNe associated with LLGRBs (filled squares), and Type II supernovae (open circles).
For reference, GRBs lie above the plot at $10^{50}\,\erg < U < 10^{52}\,\erg$, and the relativistic TDE \swift\,J1644 lies at $\approx10^{51}\,\erg$ in this framework.
For all sources, we take values of peak frequency and peak luminosity at some time (described in detail in Appendix \ref{sec:details}) and estimate velocity and energy using the same prescription that we use for AT2018cow.
Estimates of energy are sensitive to the choice of $\epsilon_B$, as illustrated with the secondary axis on the right-hand side.
}
\label{fig:vel-e}
\end{figure}

Our rederived velocity-energy diagram is shown in Figure \ref{fig:vel-e}.
AT2018cow has an energy comparable to mildly relativistic outflows (LLGRBs; e.g. SN\,1998bw)
and energetic supernovae (e.g. SN\,2007bg).
We display vertical axes for two different conventions ($\epsilon_B=1/3$ and $\epsilon_B=0.01$) to show how this affects the inferred energy.
Note that these values are not evaluated for a consistent epoch. However, for AT2018cow, we have reason to believe that the values of velocity and energy do not change significantly over the course of our observations. For other sources, it would be necessary to have a well-sampled SED over multiple epochs in order to trace the evolution of these values, and this is rare in the literature.

\subsection{A luminous millimeter transient in a dense environment}
\label{sec:luminosity}

\begin{figure}[ht]
\centering
\includegraphics[scale=1.0]{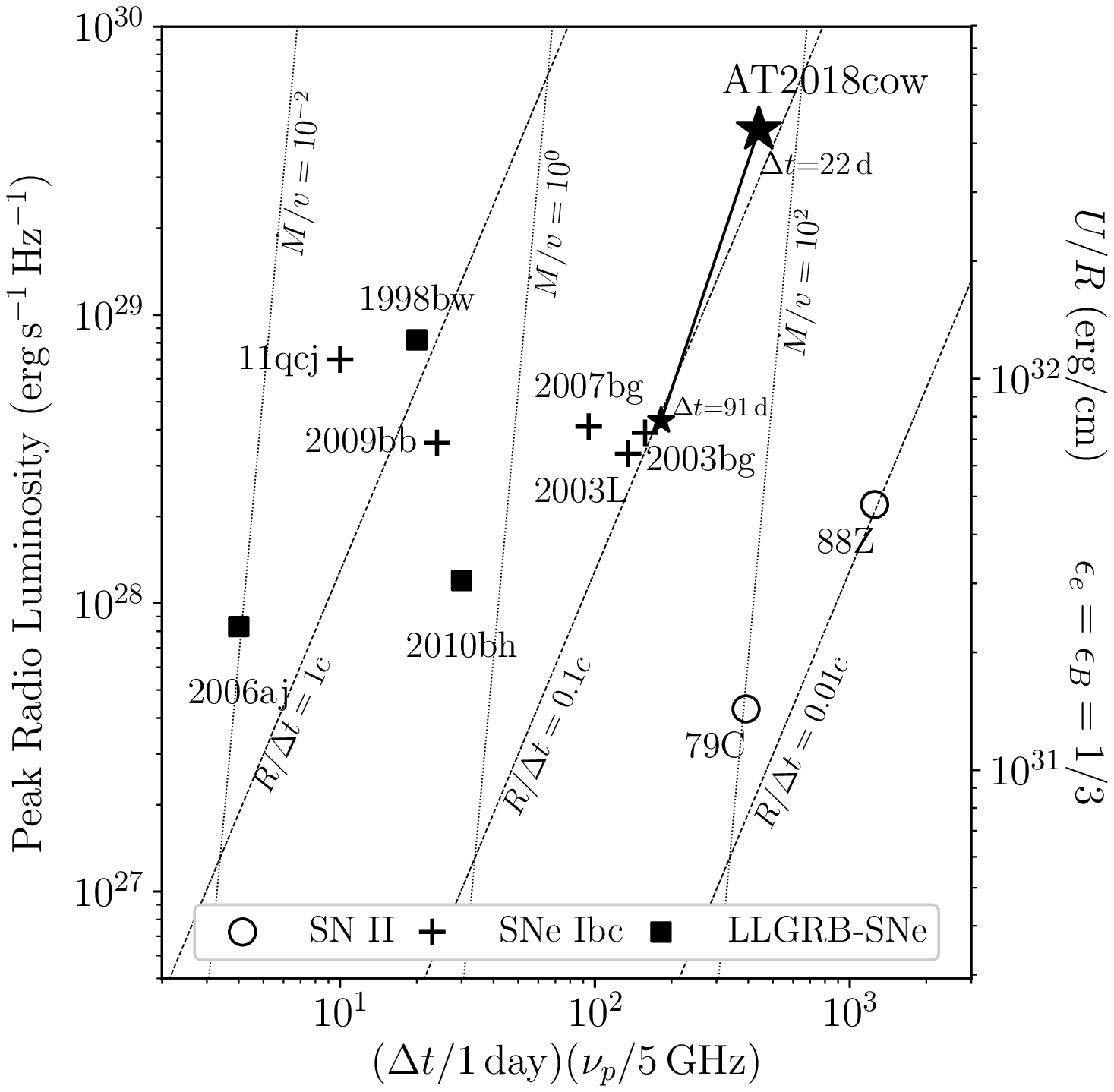}
\caption{The peak luminosity of AT2018cow on two different epochs, compared to classes of energetic transients (cf. \citealt{Chevalier1998,Soderberg2010}).
The value at $\Delta t=22\,\days$ comes from our work.
The value at $\Delta t=91\,\days$ comes from \citet{Margutti2018b}
and shows that the velocity has not slowed significantly.
For other sources, we choose values of peak frequency and peak luminosity as described in Appendix \ref{sec:details}.
AT2018cow is unusual in having a large radio luminosity as well as a high $\nu_a$,
and we discuss the physical interpretation of both of these characteristics in the text.
Lines of constant mass-loss rate (scaled to wind velocity) are shown in units of $10^{-4}\,M_\odot\,\pyr/1000\,\km\,\psec$.
Note that the dotted lines assume that the radio peak is due to synchrotron self-absorption rather than free-free absorption (FFA), but that FFA has been the preferred fit in some cases, such as for SN\,1979C and SN\,1980K \citep{Chevalier1984}}.
\label{fig:lum-tnu}
\end{figure}

Here we compare the radio luminosity of AT2018cow to that of other transients,
as observed at the spectral peak frequency at time $\Delta t$.
As shown in Figure \ref{fig:lum-tnu},
AT2018cow stands out as being several times more luminous than SN\,1998bw, and having a late peak at high frequencies.
Over time, the peak luminosity diminishes to the value reported in the low-frequency radio observations of \citet{Margutti2018b},
supporting our inference that the velocity is not changing significantly.

On this diagram we also indicate lines of constant velocity (cf. Figure 3 of \citealt{Soderberg2010} and Figure 4 of C98)
and lines of constant mass-loss rate scaled by velocity $\dot{M}/v_w$, as a diagnostic of density (cf. Figure 10 of \citealt{Jencson2018}).
Note that these lines assume $\nu_p=\nu_a$.

We now derive relations between the observational coordinates of the diagram in Figure \ref{fig:vel-e} and physical quantities: the ordinate, peak radio luminosity $L_p$, is simply a power of the energy per unit radius $U/R$.
We get an expression for $U/R$ using equation~\ref{eq:energy} and equation~\ref{eq:chev-R-short}:

\begin{equation}
    \frac{U}{R} = (3 \times 10^{29}\,\erg\,\pcm)
    \left(\frac{1}{\epsilon_B}\right)
    \left(\frac{\epsilon_e}{\epsilon_B}\right)^{-10/19} 
    \left( \frac{f}{0.5} \right)^{9/19}
    \left( \frac{L_p}{10^{26}\,\erg\,\psec\,\phz} \right)^{14/19}.
\end{equation}

This translation between $L_p$ and $U/R$ is shown on the left and right axis
labels of Figure~\ref{fig:vel-e}.

We now show that the abscissa of Figure~\ref{fig:lum-tnu}, 
$(\Delta t/1\,\days)(\nu_p/5\,\ghz)$, is very nearly proportional to the square root
of the swept up mass per unit
radius $M/R$, or equivalently, if the surrounding medium was from a pre-explosion
steady wind of speed $v_w$, $\dot{M}/v_w\propto M/R$.
A steady spherical wind of ionized hydrogen with velocity $v_w$ has 
$n_e =\dot M/(4\pi m_p r^2 v_w$), so we can reparameterize the density in terms of the mass-loss rate:

\begin{equation}
\begin{split}
    \frac{\dot{M}}{v_w} \left(\frac{1000\,\km\,\psec}{10^{-4}\,M_\odot\,\pyr}\right)
    = (0.0005)&
    \left( \frac{1}{\epsilon_B} \right)
    \left(\frac{\epsilon_e}{\epsilon_B}\right)^{-8/19} 
    \left( \frac{f}{0.5} \right)^{-26/19} 
    \left( \frac{L_p}{10^{26}\,\erg\,\psec\,\phz} \right)^{-4/19}\\
    &\times \left( \frac{\nu_p}{5\,\ghz} \right)^{2}
    \left( \frac{t_p}{1\,\days} \right)^{2}.
\end{split}
\end{equation}

Notice the weak ($-4/19$ power) dependence on $L_p$, and the quadratic dependence on 
$\nu_pt_p$, which means the lines of constant $\dot M/v_w$ are nearly vertical
in Figure~\ref{fig:lum-tnu}.

AT2018cow lies along the same velocity line as
SN\,2003bg and SN\,2003L,
but $n_e$ is a factor of a few to an order of magnitude larger.\footnote{
The radial density profile inferred for SN\,2003bg is
$n_e \approx 2.2 \times 10^{5} (r/r_0)^{-2}\,\pcmcub$ \citep{Soderberg2006}
and the radial density profile inferred for
SN\,2003L is $n_e \approx 6.1 \times 10^{4} (r/r_0)^{-2}\,\pcmcub$ \citep{Soderberg2005}.
In both cases, $r_0 \approx 10^{15}\,\cm$ is the shock radius at $t_0 = 10\,\days$. For SN\,2003L, we 
infer $t_0$ using the result that $r=3\times10^{15}\,\cm$ at $t=28\,\days$, and that $\alpha_r=0.96$ in the parameterization $r=r_0(t/t_0)^{\alpha_r}$ (Model 1 in \citealt{Soderberg2006}).}
Similarly, SN\,1998bw lies along a similar velocity line to SN\,2006aj, but $n_e$ (using our presscription)
is 40\,\pcmcub, while the density inferred for 2006aj using the same prescription is 3\,\pcmcub.
For Ibc SNe in general,
\citet{Chevalier2006} attribute the large spread in radio luminosity to a spread in circumstellar density,
with the example that SN\,2002ap (not shown in Figure \ref{fig:lum-tnu} due to its relatively low luminosity)
is roughly three orders of magnitude less luminous than SN\,2003L,
and its inferred ambient density is also a factor of three smaller.
In SN\,2003L and SN\,2003bg,
the high density was attributed to a stellar wind.

This is not the whole story: as we showed above, high peak radio luminosity just corresponds to high $U/R$, i.e. high energy and/or small radius. 
Since $U$ is the {\em converted} energy, it represents only a lower limit to the actual driving kinetic energy (becoming equal to it as the explosion transitions from free-expansion to the Sedov phase).
A higher-density medium more quickly converts the piston's energy to thermal energy than does a low density medium. Thus for a large fixed explosion energy, a denser medium will indeed lead to larger peak radio luminosities. But the direct correlation is with the (thermalized) energy per unit radius, $U/R$.  Similarly, equation~\ref{eq:chev-B-short} shows that (except for a very weak $L_p^{-2/19}$
dependence), $\nu_p\propto B_p$.  Thus higher peak frequencies are directly indicative
of a higher magnetic field, or equivalently, pressures. Thus AT2018cow's
high $\nu_p$ and high $L_p$ are quite likely mostly a consequence of it being energetic, and observed early, when the high wind density at small
radii led to high pressure, and enhanced $U/R$.  As we discuss below, this suggests
that many other supernovae could have shown similar bright mm-submm fluxes, had
they been observed at those wavelengths in their first week.

On Day 22, the inferred density is 
$\rho_0=4\times 10^{-19}\,\gram\,\pcmcub$  at a radius of $r_0=7\times 10^{15}\,{\rm cm}$ (Table 3).
From this we can infer $\dot M/v_{\rm w}=2.4\times 10^{14}\,{\rm g\,cm^{-1}}$. The mass swept up to radius $r$ is $(\dot M/v_{\rm w})r$. In Section \ref{sec:modeling} we argue that the blast wave reaches 
the edge of the surrounding bubble around $\Delta t \approx 50\,\days$.
If so, given our inferred velocity, the radius of the ``circum-bubble" is $1.7\times 10^{16}\,{\rm cm}$,
and the mass of the circum-bubble $\approx 2 \times 10^{-3}\,M_\odot$.
The mass loss rates for hot stars ($v_w\sim 2,000\,\km\,\psec$, \citealt{Lamers1993})
and red supergiants ($v_w\sim 20\,\km\,\psec$, \citealt{vanLoon2010}) range from
$10^{-4}$--$10^{-6}\,M_\odot\,\pyr$ \citep{Smith2018}. Thus $r_0/v_w$ is $\sim 1\mbox{y}$ for a hot star progenitor, and $\sim 100\mbox{y}$ for a red supergiant. Thus the circum-bubble could either have been formed by normal mass loss in a red supergiant, or end-of-life enhanced mass loss from a hot star or red supergiant (see, e.g., \citealt{Smith2017} and references therein).

UVOIR observations of AT2018cow place strong constraints on the nature of the surrounding medium \citep{Perley2018}.  The high luminosity and fast rise can be interpreted as shock-heating of a dense shell of material at $R=10^{14}\,\cm$ or 10\,\au, qualitatively consistent with the inference of dense material given the properties inferred from the radio shock.  On the other hand, early spectra show no narrow emission lines indicative of a shock and the light curve declines steeply after peak, both of which suggest that this dense material must also be quite limited in extent, with little material at larger radii.  While the radio observations also suggest a cutoff in the density distribution may exist, the $0.1c$ shock does not reach it for almost 20 days, a quite different timescale than the optical peak (reached in less than 3 days) or early spectroscopy.  This might be due to the $0.1c$ shock being produced by breakout from the $R=10^{14}\mbox{cm}$ shell which re-energized the (much slower) supernova shock. Or there could be deviations from spherical asymmetry (for example, with the optical heating a quasi-spherical shell but the radio shock passing through a denser toroidal component or clouds along a bipolar jet).

\begin{figure}[ht]
\centering
\includegraphics[scale=0.65]{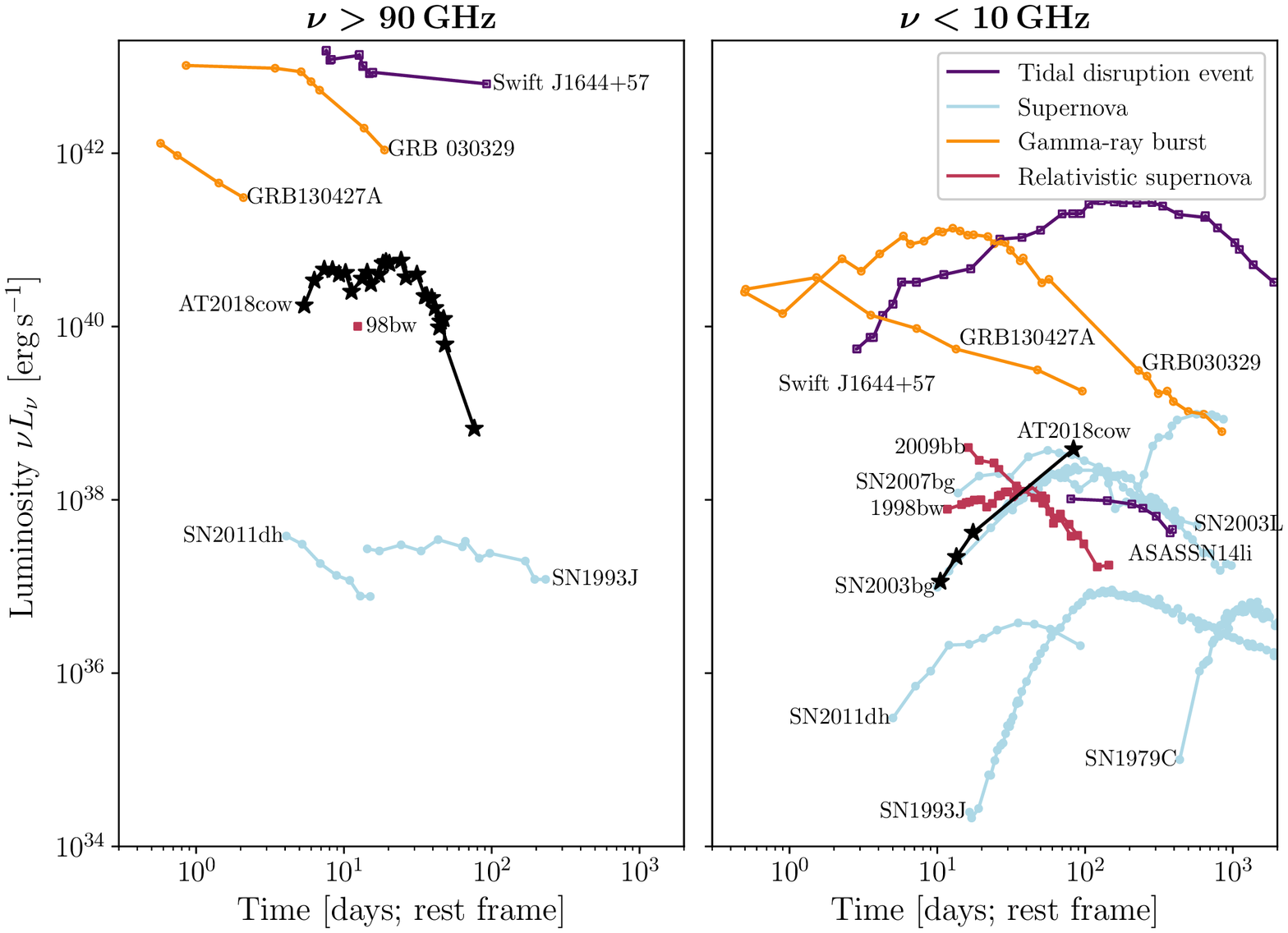}
\caption{Luminosity evolution for different transients, measured at high frequencies ($\nu > 90\,\ghz$; left panel) and low frequencies ($\nu < 10\,\ghz$; right panel).
Classes are GRBs (orange open circles; \citealt{Sheth2003,Berger2003,Perley2014}), TDEs (purple open squares; \citealt{Zauderer2011,Berger2012,Zauderer2013,Alexander2016,Eftekhari2018}), non-relativistic supernovae (light blue filled circles; \citealt{Weiler1986,Weiler2007,Soderberg2005,Soderberg2006,Salas2013,Horesh2013,Krauss2012}), and relativistic supernovae (red filled squares; \citealt{Kulkarni1998,Soderberg2010}).
Thus there are a number of transients measured with radio telescopes (relativistic SN\,2009bb, energetic supernovae 2003L, 2003bg, and 2007bg) that could have been bright millimeter transients but were not observed at high frequencies.
The late-time low-frequency AT2018cow point is from \citet{Margutti2018b}.
}
\label{fig:mm-tau-lum}
\end{figure}

Thus an energetic shock propagating into a dense environment
could produce a radio SED that peaks at sub-millimeter
wavelengths at early times.
However, as illustrated in the left panel of Figure \ref{fig:mm-tau-lum}
searches at high frequencies at early times have been rare,
and primarily limited to
transients with relativistic jetted outflows (GRBs, TDEs).
We suggest that these searches be expanded to
other classes of transients:
luminous SNe such as SN\,2003L and SN\,2007bg,
and luminous TDEs such as ASASSN14li,
all exploded into dense media and exhibited luminous centimeter-wavelength emission at $t>10\,\days$.
As time goes on, the SED peak shifts to lower frequencies and diminishes in brightness,
so these events could have been bright millimeter transients at $t < 10\,\days$.
This is supported by Figure \ref{fig:lum-tnu}, which shows that 
SN\,2007bg, SN\,2003bg, and SN\,2003L could have appeared similar to AT2018cow had they been observed earlier at higher frequencies.

\subsection{Novel features of the synchrotron model parameters}
\label{sec:breakfreq}

The ordering of the break frequencies,
$\nu_\mathrm{ff} < \nu_m < \nu_c < \nu_a$,
is an unusual regime for long-wavelength observations.
For a relativistic shock (GRBs), the typical orderings are $\nu_a < \nu_c < \nu_m$ (the fast cooling regime) and $\nu_a < \nu_m < \nu_c$ (the slow cooling regime; \citealt{Sari1998}).
For non-relativistic shocks, the ordering in most cases seems to be $\nu_a < \nu_m < \nu_c$ at measured frequencies above 1.4\,GHz, but can also be $\nu_m < \nu_a < \nu_c$; $\nu_c$ is typically considered unimportant for long-wavelength observations \citep{Nakar2011}. 

The low cooling frequency is a consequence of a large magnetic field strength, $\nu_c \propto B^{-3}$ (reduced even further for $t<10\mbox{d}$ by Compton cooling on the UVOIR flux, which dominates over synchrotron cooling).
This in turn presumably arises from the injection of a large amount of energy into a small volume of material,
consistent with the low velocity we measure.
From equation~\ref{eq:chev-B-short} we see that $B_p$ scales as $(\epsilon_e/\epsilon_B)^{-4/19} L_p^{-2/19} \nu_p$.
Changing $\epsilon_B$ from 1/3 to 0.01 could increase $\nu_c$ by a factor of 8, still much lower than our observed frequencies.
This regime is selectively probed by sub-millimeter observations,
because a low $\nu_c$ (high $B_p$) gives rise to a $\nu_a$
that falls in this wavelength regime.
We note that in the same framework,
the relativistic TDE \swift\,J1644+57 (whose long-wavelength SED also peaked in the sub-mm for the first few weeks) 
would also have had a cooling frequency 
below much of the observed frequency range ($\nu_c \approx 6\,\ghz$).

Since
$\nu_c$ is below any of our measured frequencies,
the injection spectrum (the spectrum of the electrons prior to cooling) has a shallower power-law index than what we measure, $p_i=p-1$.
This suggests that $p_i\approx2.2$ on Day 22, when $p\approx3.2$,
This is not unreasonable for Fermi acceleration from a strong shock. Typical young Galactic supernova remnants have $p=p_i=2.4$
in the radio, flattening to $p\sim 2$ at higher frequencies
\citep{Urosevic2014}.


\begin{figure}[ht]
\centering
\includegraphics[scale=0.85]{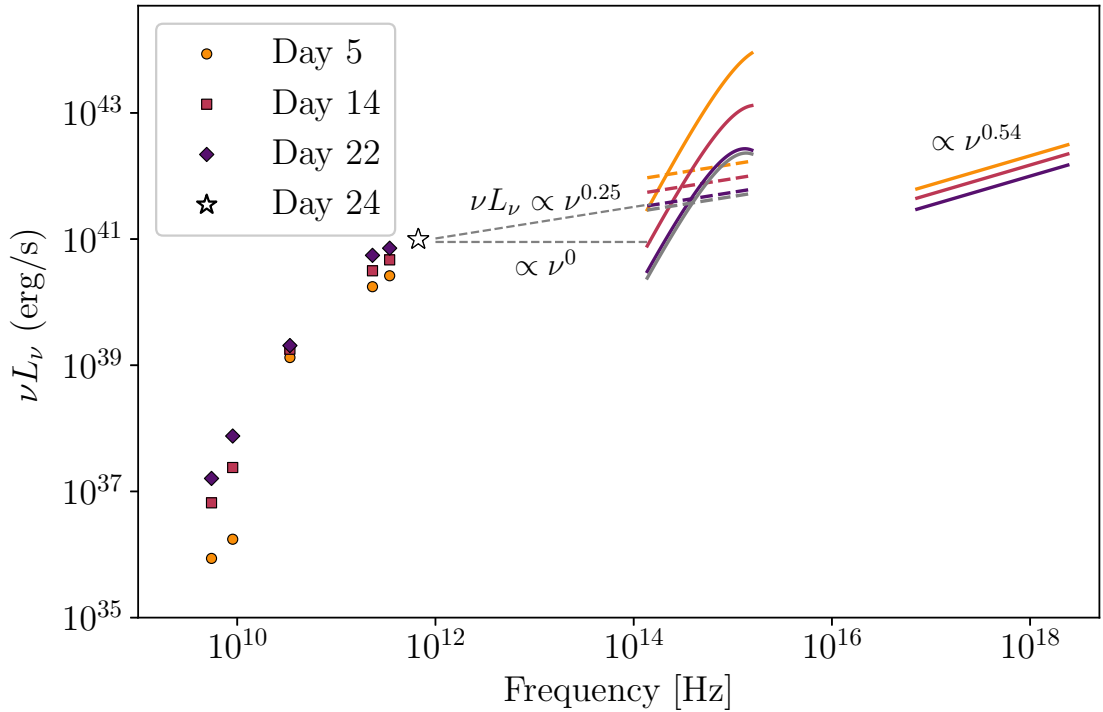}
\caption{The full radio to X-ray SED. Since the ATCA data vary smoothly over the course of our observations, we fit a power law to the existing light curves (see Figure~\ref{fig:lc}) and plot the values for the given day at 5.5\,\ghz, 9\,\ghz, and 34\,\ghz. For the SMA data, we interpolate the spectrum for the given day and plot the value at 231.5\,\ghz\ and 345\,\ghz.
We plot the ALMA data as measured,
including the single Band 9 measurement (white star) which seems to show an excess above the other radio data.
We plot the best-fit blackbody and nonthermal component
from \citet{Perley2018},
and show that the nonthermal component could be an extension of the excess seen in Band 9 on Day 24.
We plot the \swift/XRT data as follows:
we interpolate the light curves to estimate the integrated 0.3--10\,\kev\ flux at the given epoch.
We use the geometric mean of (0.3\,\kev, 10\,\kev) and the spectral index $\nu^{0.54}$ to solve for the normalization coefficient for the spectrum.
We display the spectrum across the full XRT range.
}
\label{fig:sed}
\end{figure}

\section{Origin of the X-ray emission and emergence of a compact source}
\label{sec:xray}

During the plateau phase,
the fluence in the \swift/XRT bands
is $\int F_Xdt \approx 1.7\times 10^{-5}\,{\rm erg\,cm^{-2}}$.
Integrating the \swift/XRT light curve until $\Delta t=22\,\days$, we find $7 \times 10^{48}\,\erg$.
Using the \nustar\ spectra index $\alpha=0.5$,
extrapolating this to $100\,\kev$ would increase this energy by a factor of three.

The total X-ray energy emitted in the first three weeks is thus
greater than the total energy in the shock inferred from radio observations on Day 22, $U  \approx 4 \times 10^{48}\,\erg$.
If a significant proportion of the X-rays are produced by IC emission,
then our assumption of $\epsilon_B=\epsilon_e=1/3$
clearly results in an underestimate of the total energy,
and an assumption of $\epsilon_B=0.01$ would be more appropriate.
As shown in Figure~\ref{fig:vel-e},
$U$ would be increased by a factor of 9 with the assumption
$\epsilon_B=0.01$ and $\epsilon_e=0.1$, just barely
comparable to the total energy emitted in X-rays.

The luminosity of the UV/optical/IR (UVOIR) source declines $\propto t^{-\beta}$, where $\beta\approx 2.5$ \citep{Perley2018}.
Assuming a constant expansion speed for the shock of $0.13c$ (see Table \ref{tab:measurements} and
related discussion in Section~\ref{sec:modeling}) the photon energy density of the UVOIR source,
$u_{\rm ph}=L_{\rm UVOIR}(T)/(c4\pi R(t)^2)
= 5.2\,(t/10\,\days)^{-9/2}(\epsilon_e/\epsilon_B)^{2/19} \,{\rm erg\,cm^{-3}}$.
Assuming that the magnetic field pressure scales with the ram pressure of the shock ($\rho_1 v_1^2$) and assuming $\rho \propto r^{-2}$, from equation~\ref{eq:chev-B-short}, the magnetic energy density
$u_B = 7.4(t/10\,\days)^{-2}(\epsilon_B/\epsilon_e)^{8/19}\,\erg\,\pcmcub$.
For $\epsilon_B=0.01$, $\epsilon_e=0.1$, $u_\mathrm{ph}/u_B=2(t/10\mbox{d})^{-2.5}$, with only a rather weak dependence on the epsilons.
This ratio is equal to unity around $t=13\,\days$, marking the transition from a regime dominated by Compton cooling to a regime dominated by synchrotron cooling.

This ratio is much lower than the observed ratio $L_X/L_{\mathrm{radio}}\gtrsim 30$, and the X-ray spectral index is also substantially flatter than the radio spectral index.
We conclude that the X-ray emission during the plateau phase does not naturally arise from IC scattering of the UVOIR source by the electrons in the post-shocked region (which also generate the radio to sub-millimeter emission via synchrotron radiation):
IC from the radio-mm emitting region alone underpredicts the X-ray luminosity, predicts an X-ray luminosity declining much more rapidly than observed
in the first 20 days, and predicts too steep a spectrum.
It also does not naturally arise from an extension of the $\alpha\approx -1.1$ radio-submm synchrotron spectrum: the X-ray emission is
some 25 times brighter than that extrapolation (see Figure~\ref{fig:sed}), and has a much flatter ($\alpha\approx -0.5$) spectral index.  Further speculative 
modelling of the source of the X-ray emission during the plateau phase is beyond
the scope of this paper.

During the decline phase $t>20\,\days$, the timescale
of these fluctuations is around $0.05t$,
while the diameter
we infer for the radio-emitting region (see Section
\ref{sec:modeling}) is $\sim 2\times 0.13t=0.26t$.
Thus the X-ray emission
must arise in a different and more compact source than the radio-emitting shell.

From the plateau phase to the decline phase,
the X-ray emission softens, as shown
in the bottom panel of Figure~\ref{fig:lc}
and reported by \citet{Kuin2018}.
From the \nustar\ data,
measuring the flux using \texttt{cflux},
we infer a hardness ratio $L_X$(10--200\,\kev)/$L_X$(0.3-10\,\kev)$\approx26$ on Epoch 1,
similar to what is inferred by \citet{Kuin2018} using a joint BAT/XRT analysis.
From the \nustar\ data, we find a hardness ratio of $L_X$(10--200\,\kev)/$L_X$(0.3--10\,\kev)$\approx$4--5 on Epochs 3 and 4.
This is consistent with other studies,
which found negligible spectral evolution in the \swift\ 0.3--10\,\kev\ band, but significant spectral evolution at higher energies \citep{Kuin2018,RiveraSandoval2018d,Margutti2018b}.

Thus these two changes, the onset of variability 5 times faster than the light-travel
time across the radio-emitting shell, and the striking change in the
spectrum, lead us to conclude that the beyond 20\,\days\ the X-ray emission 
arises from a different and more compact source than during the plateau phase--
in the decline phase we are, arguably,
probing regions closer to the central engine of the event.

The peculiarities of the UVOIR spectrum have led some to propose that AT2018cow is a tidal disruption event (TDE) of a white dwarf by a $\sim 10^{5-6}M_\odot$ black hole \citep{Perley2018,Kuin2018}. 
Given its off-nucleus location (1.7\,\kpc; \citealt{Perley2018})
in a star-forming galaxy, and the similarities of the radio-emitting shock to those of other supernovae, it seems more natural to suppose that
AT2018cow originated in a stellar cataclysm. 
Ultimately, however, our radio observations only require a $v\sim 0.1c$ shock wave propagating into a dense medium, which could very plausibly arise in both TDE and supernova models. The radio observations do little to distinguish them.
In either picture, the striking late-time change in the X-ray behavior suggests the emergence of a central engine.
In the TDE case, this could be an accretion disk around a black hole.
In the stellar explosion case,
this could be a natal black hole accreting (fall-back) matter from the debris, or a magnetar.
The emergence could then be due to
a channel between the interior and the surface opened up by a collimated outflow (a ``jet'' or stifled jet's cocoon breakout; \citealt{Nakar2015}),
or to gaps in the photosphere opened by Rayleigh-Taylor instabilities.

We now briefly explore the magnetar model,
which has been proposed for cosmological long-duration gamma-ray bursts (e.g. \citealt{Thompson2004}) and superluminous supernovae (e.g. \citealt{Kasen2010}).
\citet{Prentice2018} invoked a magnetar model to explain the UVOIR observations of AT2018cow and found a best-fit magnetic field strength of $2 \times 10^{15}\,\gauss$ and a best-fit spin period of $11\,\mathrm{ms}$.
Note that the magnetar itself need not be directly visible: the X-rays we see could be due to the emergence of bubbles of the magnetar-powered wind nebula
\citep{Kasen2016}. The spin-down luminosity of a magnetar with period $P$ is $L\propto \omega\dot{\omega}$
where $\omega=2\pi/P$ is the angular frequency. The spin-down timescale is
$\tau_c=P/2\dot P$. 
We set $\tau_c=20\,\days$, $L_X=5 \times 10^{42}\,\erg\,\psec$ and find $P = 50\,\millisecond$ and $\dot{P}=1.4 \times 10^{-8}\,\second\,\psec$, assuming that
all the spin-down power goes into X-ray production. 
For a constant spin-down rate, this would correspond to an initial spin period of 26\,\millisecond, similar to the result in \citet{Prentice2018} for the model fit to the \emph{griz} light curve.

With $P$ and $\dot{P}$ in hand, using the standard dipole formula, we find a lower limit on the
magnetic field strength of $8\times 10^{14}\,$G,
which is consistent with the value found in \citet{Prentice2018}.
Our modeling of the forward shock led to a lower limit to the
energy of $U\approx 10^{49}\,$erg, depending on the value of $\epsilon_B$. If this was supplied by a magnetar then 
the initial period of the magnetar is $\lesssim 10\,\mathrm{ms}\,(U/10^{50}\,{\rm erg})^{-1/2}$. We end this discussion by noting that the spin-down luminosity in the dipole model (with constant B-field) is $\propto t^{-2}$,
which is roughly consistent with the
slope of the decay of the X-ray light curve.
This is, however, not easily distinguished from the
$\propto t^{-5/3}$ slope expected from accretion in a TDE \citep{Phinney1989} or fallback \citep{Michel1988}. 

If this is a stellar explosion,
then the features in the UVOIR spectra and the rise time
point to an extended progenitor ($10^{14}\,\cm$; \citealt{Perley2018}), comparable in size to the largest red supergiants.
This is not consistent with the compact, stripped stars invoked as progenitors for other classes of engine-driven explosions like GRBs and SLSNe (although see \citet{Smith2012} for a possible exception).
As discussed in \citet{Perley2018} and Section \ref{sec:luminosity}, a more likely scenario is that 
the progenitor experienced a dramatic, abrupt episode of mass loss shortly before the explosion, and the UVOIR photosphere lies within this ``brick wall'' which the supernova blast wave struck and re-thermalized. 

Regardless of the nature of the central engine,
we have the following model:
the fastest-moving ejecta races ahead at $v_1= 0.13c$ into a dense ``circum-bubble" of radius, $R_b$. 
In the post-shocked gas electrons are accelerated into a power-law spectrum and magnetic fields are amplified.
We attribute decay of the resulting radio emission at $t_b=50\,\,\days$
to the fast-moving ejecta reaching the edge of this circum-bubble and infer a radius $R_b=v_1t_b\approx 1.7\times 10^{16}\,{\rm cm}$,
and a mass of $10^{-3}\,M_\odot$.
Within the radio-emitting shell is a long-lived engine which
may inflate a bubble of plasma and magnetic fields \citep{Bucciantini2007}.
There is also slower ejecta heated by a central source (or radioactivity) that expands and emits UVOIR radiation.
The photosphere of this component recedes with time,
and at early times its large Compton optical depth obscures direct emission from the vicinity of the central engine.
At later times, this central region emerges.

\section{Conclusions and Outlook}
\label{sec:outlook}

Persuasive arguments can and have been made for both supernova and tidal disruption event origins for AT2018cow.
Our extensive radio through sub-millimeter observations
enable us to draw definitive conclusions about the outer blast wave launched following the event,
independent of the origin of the event.
The blast wave is sub-relativistic ($v/c=0.13$)
and plows into a dense medium ($n_e=3\times 10^5\,{\rm cm}^{-3}$ at $R\sim 7\times 10^{15}\mbox{cm}$).
The energy contained within this blast wave is $U\approx 10^{49\pm 0.3}\,\erg$.
In contrast to the UVOIR luminosity, which declines as $L_{UVOIR}\propto t^{-2.5}$
over the period $3\mbox{d}<t<20\mbox{d}$,
the mm-submm and X-ray luminosities 
are both relatively constant over the same period (with $\sim 50$\%\ variations over timescales of a few days, comparable to the
light-travel time across the radio-emitting shell, $t_{lc}=2vt/c=0.26t$; see Figure~\ref{fig:lc}). 

The initially attractive idea of attributing the X-rays to IC scattering of the rapidly declining flux of 
UVOIR photons by relativistic electrons (which give rise to the radio and sub-millimeter flux) is not naturally consistent with the slow
decline of the X-ray and radio-submm flux, the X-ray luminosity, or the X-ray spectral index.
Thus we are forced to invoke an additional source of X-ray emission during both
the $t<20\mbox{d}$ plateau phase, and during the decline phase
after 20 days, when the X-rays begin to fade and 
show dramatic variations now on timescales several times {\em shorter} than the
light-travel time across the radio-emitting shell.
These are suggestive of power from a central engine, which could be consistent with either a stellar explosion or a TDE.
Future X-ray monitoring may be useful in differentiating between central engine models;
in particular, a power-law decay ($t^{-2}$) is a distinct signature of the magnetar model, although difficult in practice to distinguish from the $t^{-5/3}$ expected from fall-back or a tidal disruption event.

The radio source is remarkable even on purely observational grounds.
The peak radio luminosity (nearly $10^{41}\,\erg\,\psec$)
greatly exceeds that of the most radio-luminous supernovae and `normal' TDEs, and is surpassed only by relativistic jetted transients (GRBs and TDEs).
The source remains luminous at sub-millimeter wavelengths
for nearly a month,
with a self-absorption frequency $\nu_a\sim 100\,\ghz$
at $\Delta t\approx22\,\days$.

The source is strongly detected at nearly a terahertz (ALMA Band 9; 671\,\ghz) even three weeks post discovery.
We note that the Band 9 flux is higher than the extrapolation
based from lower frequency bands (Figure~\ref{fig:spec}, middle panel),
and intriguingly connects to the NIR non-thermal component
suggested by \citet{Perley2018}. However, we readily admit that 
the apparent excess in the Band 9 flux is only 2$\sigma$, and also note that the case for the NIR non-thermal component
is not secure.

Finally, it is worth re-iterating that AT2018cow is a mere 60\,Mpc away.
The proximity hints at an extensive population of which AT2018cow is the prototype.
The key distinction between
AT2018cow and other fast transients is the strong millimeter and
sub-millimeter emission.
In general, it is apparent from Figure~\ref{fig:lum-tnu}
that an energetic shock propagating into a dense medium will exhibit strong
millimeter emission during the first weeks. Many other supernovae would likely 
have had bright emission at mm-submm wavelengths, had they been observed
early at those wavelengths. Combining velocities measured at very early times at
such short wavelengths, with much later observations at low frequencies
could reveal the slowing of the shock associated with the
transition from free expansion to the Sedov phase, constraining the total
energy in relativistic ejecta.
Taken together, these two developments, given
that we are now squarely in the era of industrial optical time domain astronomy
(e.g.\ PS-1, PS-2, ASAS-SN, ATLAS, ZTF and soon BlackGEM), argue
for a high-frequency facility dedicated to the pursuit of transients.

The code used to produce the results described in this paper was written in Python
and is available online in an open-source repository\footnote{https://github.com/annayqho/AT2018cow}.

\appendix

\section{ALMA Band 9 Calibration}
\label{sec:appendix-band9}

\begin{figure}[ht]
\centering
\includegraphics[scale=0.5]{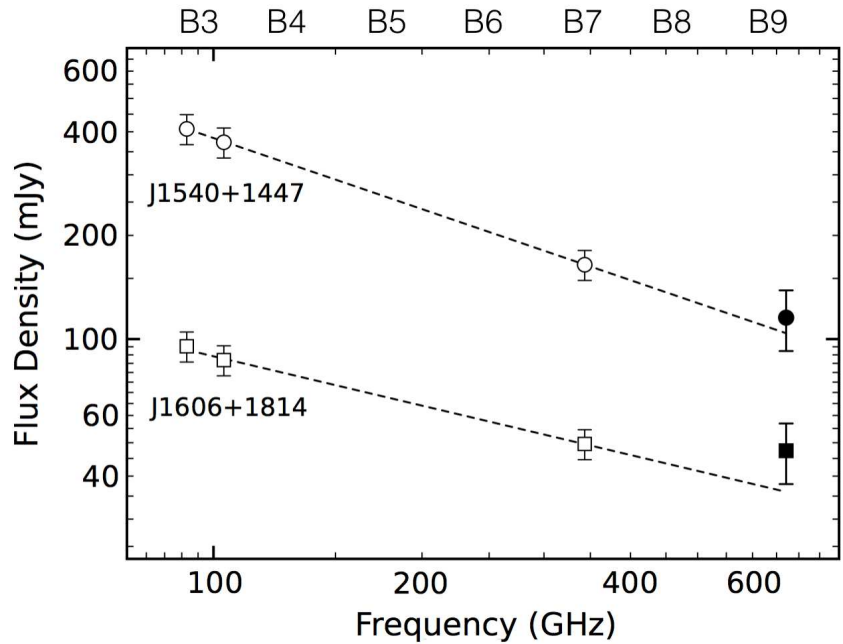}
\caption{ALMA flux measurements of secondary calibrators J1540+1447 and J1606+1814, validating the absolute flux calibration of the Band 9 data. The solid symbol shows our ALMA Band 9 measurement on 10 July 2018. The open symbols represent the estimated flux densities in Bands 3 and 7 on July 10, derived by interpolating between archival values measured by ALMA on 18 May and 9 Aug 2018. Error bars represent 10$\%$ uncertainties for the archival data and 20$\%$ for our measured Band 9 data. The straight dashed lines have been added for purpose of visualization and do not represent an actual fit to the data. ALMA bands B3--B9 are shown at the top for reference.}
\label{fig:calibrators}
\end{figure}

For ALMA Band 9, due to the relatively low signal-to-noise of the data, all eight spectral windows were combined to derive a combined phase solution, which was then mapped to each individual spectral window. As a result, no in-band analysis of the spectral index was done and a single flux value was derived for the Band 9 imaging. When fitting a Gaussian function to the Band 9 image, the source is represented as a point-source, suggesting that the image is of good enough quality to derive a meaningful flux densities. A phase-only self-calibration did not provide good solutions, it decreased the phase coherence, and it resulted in an image that could no longer be fitted with a point-source. Therefore, we did not apply a self-calibration to the Band 9 data. To verify that changing weather conditions did not affect the phase coherence in the data, we split the data in three different time-bins and imaged each time-bin separately. The change in flux density between the different time-bins was within 6$\%$. Similarly, imaging the data from the short, intermediate and long baselines by splitting the data in three bins in uv-range (12$-$90m, 90$-$170m and 170$-$312m) showed a difference in flux density $<$13$\%$, despite the sparser antenna distribution and poor uv-coverage in the long-baseline bin. Also, the XX and YY polarization images were similar to within 3$\%$. To examine the reliability of the absolute flux calibration of the Band 9 data, we imaged the two secondary calibrators, the quasars J1540+1447 and J1606+1814, using the same flux and bandpass calibrator as for AT2018cow. Figure~\ref{fig:calibrators} shows that the Band 9 flux densities of these two secondary calibrators are in reasonable agreement with values from the ALMA calibrator catalog in the lower bands if there is no spectral curvature, although uncertainties in absolute flux calibration may have led us to slightly overpredict our derived Band 9 values. In all, our tests are consistent with the ALMA Band 9 flux density being accurate to within a 20$\%$ uncertainty, which is standard for high-frequency ALMA observations.

\section{Full SMA light curves}
\label{sec:appendix-smalc}

\begin{figure}[p]
\centering
\includegraphics[scale=0.8]{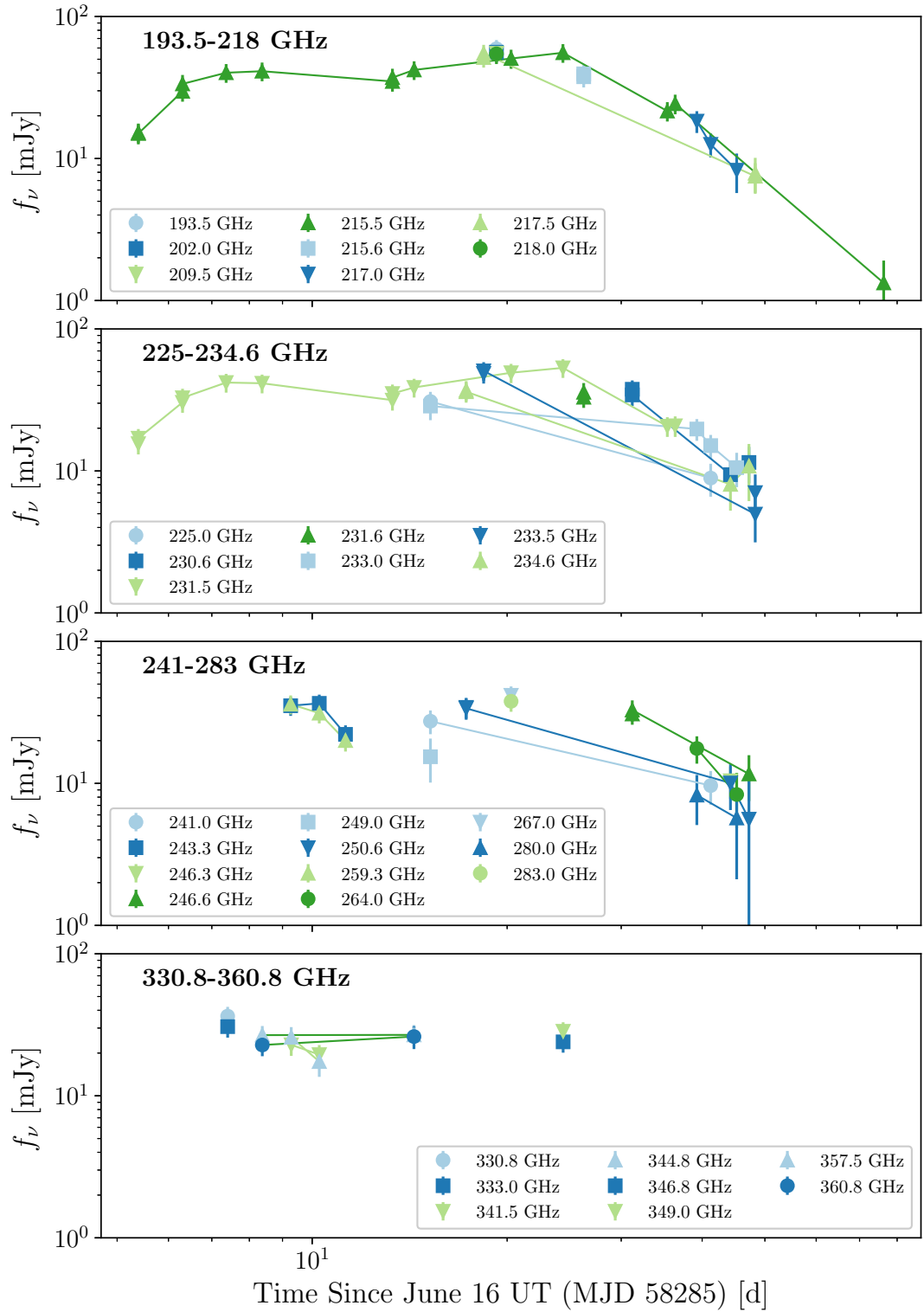}
\caption{Full SMA light curves of AT2018cow for each individual frequency tuning}
\label{fig:full-lc}
\end{figure}

Figure~\ref{fig:full-lc} shows the full
set of SMA light curves, grouped by frequency.

\section{Selection of peak frequency and peak luminosity for other transients}
\label{sec:details}

\begin{itemize}

    \item \textbf{GRB\,130427A} ($z=0.340$): For Figure~\ref{fig:mm-tau-lum}, we use the 93\,\ghz\ light curve and 5.1\,\ghz\ light curve from \citet{Perley2014}.

    \item \textbf{GRB\,030329} ($z=0.1686$): For Figure~\ref{fig:mm-tau-lum}, we use the 250\,\ghz\ light curve at high frequencies \citep{Sheth2003}, and at low frequencies the 8.5\,\ghz\ and 2.3\,\ghz\ light curves from \citet{Berger2003} and \citet{vanderHorst2008} respectively.

    \item \textbf{SN\,2009bb} ($d=40\,\mpc$): 
    For Figures \ref{fig:vel-e} and \ref{fig:lum-tnu},
    we use the report in \citet{Soderberg2010} that from their first spectrum at $\Delta t = 20\,\days$, they infer $\nu_p=6\,\ghz$ and $L_p \approx 3.6 \times 10^{28}\,\erg\,\psec\,\phz$.
    For Figure~\ref{fig:mm-tau-lum}, we use the 8.5\,\ghz\ light curve from \citet{Soderberg2010}.

    \item \textbf{SN\,1998bw} ($d=38\,\mpc$): 
    For Figures \ref{fig:vel-e} and \ref{fig:lum-tnu},
    we use the report in \citet{Kulkarni1998} that on Day 10 the peak flux is 50\,\mjy\ at 10\,\ghz.
    For Figure~\ref{fig:mm-tau-lum}, we show the single 150\,\ghz\ measurement by SCUBA and the 2.3\,\ghz\ light curve from \citet{Kulkarni1998}.

    \item \textbf{SN\,2006aj} ($z=0.03345$): 
    For Figures \ref{fig:vel-e} and \ref{fig:lum-tnu}, 
    we use the report in \citet{Soderberg2006} that at 5\,\days\ the radio spectrum peaks near 4\,\ghz. They do not report the peak luminosity, so we use the reported flux of 4.86\,\ghz\ at 5\,\days, which is 328\,\ujy.

    \item \textbf{SN\,2010bh} ($z=0.0593$): 
    For Figures \ref{fig:vel-e} and \ref{fig:lum-tnu}, 
    we use the report in \citet{Margutti2013} that at 30\,\days, $\nu_a \approx 5\,\ghz$ and $F_{\nu,a} \approx 130\,\ujy$.

    \item \textbf{PTF\,11qcj} ($z=0.0287$): 
    For Figures \ref{fig:vel-e} and \ref{fig:lum-tnu},
    we use the report in \citet{Corsi2014} that the peak luminosity at 5\,\ghz\ was $7 \times 10^{28}\,\erg\,\psec\,\phz$ at 10\,\days.

    \item \textbf{SN\,2011dh} ($d=8.03\,\mpc$): For Figure~\ref{fig:mm-tau-lum}, we use the 107\,\ghz\ and 93\,\ghz\ light curves at high frequencies \citep{Horesh2013} and the 8.5\,\ghz\ and 6.7\,\ghz\ light curves at low frequencies \citep{Horesh2013,Krauss2012}.
    
    \item \textbf{SN\,2007bg} ($d = 152\,\mpc$): 
    For Figures \ref{fig:vel-e} and \ref{fig:lum-tnu},
    we use the report in \citet{Salas2013}
    that in Phase 1 of the explosion,
    the peak luminosity was $4.1 \times 10^{28}\,\erg\,\psec\,\phz$ at 8.46\,\ghz\ on Day 55.9.
    For Figure~\ref{fig:mm-tau-lum}, we use the 8.5\,\ghz\ light curve from \citet{Salas2013}.

    \item \textbf{SN\,2003L} ($d=92\,\mpc$): 
    For Figures \ref{fig:vel-e} and \ref{fig:lum-tnu},
    we use the report in \citet{Soderberg2005} that at 30\,days, the peak flux density was 3.2\,mJy at 22.5\,\ghz. For Figure~\ref{fig:mm-tau-lum}, we use the 8.5\,\ghz\ light curve because it is the best-sampled over the largest range of time.

    \item \textbf{SN\,2003bg} ($d=19.6\,\mpc$):  
    For Figures \ref{fig:vel-e} and \ref{fig:lum-tnu},
    we use the report in \citet{Soderberg2006} that the peak flux density is 85\,\mjy\ at 22.5\,\ghz\ on Day 35.
    For Figure~\ref{fig:mm-tau-lum}, we use the 8.5\,\ghz\ light curve from \citet{Soderberg2006}.
    
    \item \textbf{SN\,1993J} ($d=3.63\,\mpc$): For Figure~\ref{fig:mm-tau-lum},
    we use the 5\,\ghz\ light curve at low frequencies and the 99.4\,\ghz\ light curve at high frequencies \citep{Weiler2007}.

    \item \textbf{SN\,1988Z} ($z=0.022$): 
    For Figures \ref{fig:vel-e} and \ref{fig:lum-tnu}, we use the report in \citet{vanDyk1993} that the 6\,\cm\ maximum flux density was 1.90\,\mjy, at 1253\,\days\ after the explosion.
    
    \item \textbf{SN\,1979C} ($20\,\mpc$):  
    For Figure~\ref{fig:mm-tau-lum},
    we use the 1.4\,\ghz\ light curve at low frequencies and the 99.4\,\ghz\ light curve at high frequencies \citep{Weiler1986,Weiler1991}.
    For Figures \ref{fig:vel-e} and \ref{fig:lum-tnu},
    we simply use the peak of the 1.4\,\ghz\ light curve, which is roughly 12\,\mjy\ at 1400\,\days.
    
    \item \textbf{\swift\,J1644+57} ($z=0.354$):
    For Figures \ref{fig:vel-e} and \ref{fig:lum-tnu},
    we use the reported $\nu_p, F_p$ on Day 15 (corrected to Day 18 in \citet{Eftekhari2018}).
    For Figure~\ref{fig:mm-tau-lum}, we use the 225\,\ghz\ and 230\,\ghz\ light curves from the SMA \citep{Zauderer2011,Berger2012}, adding 3.04 days to the \citet{Zauderer2011} points because (as described in \citealt{Eftekhari2018}) subsequent analysis of the BAT data revealed emission earlier than had been previously noticed.
    We use 4.9\,\ghz\ data from \citet{Berger2012}, \citet{Zauderer2013},
    and \citet{Eftekhari2018}.
    
\end{itemize}

\acknowledgements

The authors are grateful to the staff at the SMA, the CSIRO Astronomy and Space Science (CASS), and ALMA for rapidly scheduling and executing the observations and reducing the data.
It is a pleasure to thank John Carpenter for his guidance and his assistance with the ALMA observations.
Thank you to Dale Frail, Raffaella Margutti, and Roger Chevalier for providing feedback on the manuscript,
and Gregg Hallinan, Dillon Dong, and Jacob Jencson for helpful discussions.
Finally, we would like to thank the anonymous referee for thoughtful suggestions that greatly improved the clarity of the paper.

A.Y.Q.H. was supported by a National Science Foundation Graduate Research Fellowship under Grant No. DGE‐1144469.
This work was supported by the GROWTH project funded by the National Science Foundation under PIRE Grant No 1545949.
This research was funded in part by the Gordon and Betty Moore Foundation
through Grant GBMF5076 to ESP, and A.Y.Q.H., E.S.P. and S.R.K. benefited from
interactions with Dan Kasen, David Khatami and Eliot 
Quataert funded by that grant.
TM acknowledges the support of the Australian Research Council through grant FT150100099.
DD is supported by an Australian Government Research Training Program Scholarship.

The Submillimeter Array is a joint project between the Smithsonian Astrophysical Observatory and the Academia Sinica Institute of Astronomy and Astrophysics and is funded by the Smithsonian Institution and the Academia Sinica.
The Australia Telescope Compact Array is part of the Australia Telescope National Facility which is funded by the Australian Government for operation as a National Facility managed by CSIRO.
This paper makes use of the following ALMA data: ADS/JAO.ALMA\#2017.A.00047.T. ALMA 
is a partnership of ESO (representing its member states), NSF (USA) and NINS (Japan), 
together with NRC (Canada), MOST and ASIAA (Taiwan), and KASI (Republic of Korea), in 
cooperation with the Republic of Chile. The Joint ALMA Observatory is operated by 
ESO, AUI/NRAO and NAOJ.
The National Radio Astronomy Observatory is a facility of the National Science Foundation operated under cooperative agreement by Associated Universities, Inc.
This work made use of data supplied by the UK Swift Science Data Centre at the University of Leicester.

\software{
\code{Astropy} \citep{Astropy2013, Astropy2018},
\code{IPython} \citep{ipython},
\code{matplotlib} \citep{matplotlib},
\code{numpy} \citep{numpy},
\code{scipy} \citep{scipy}}

\end{document}